\documentclass[aps,prb,reprint,superscriptaddress,amsmath,amssymb,nofootinbib]{revtex4-2}
\usepackage{graphicx}
\usepackage{microtype}
\usepackage{tikz}
\definecolor{HFbar}{HTML}{006B69}
\definecolor{HFact}{HTML}{254D83}
\definecolor{HFcut}{HTML}{9E4022}
\definecolor{HFgray}{HTML}{71808A}
\usepackage[hidelinks]{hyperref}
\newcommand{\ket}[1]{\lvert #1\rangle}
\newcommand{\bra}[1]{\langle #1\rvert}
\newcommand{\id}{\mathbf 1}
\DeclareMathOperator{\tr}{tr}
\hypersetup{pdftitle={Inequivalent Hilbert space fragmentations from locally similar Hamiltonians},
  pdfauthor={Somnath Maity, Pramod Padmanabhan, Vladimir Korepin}}

\begin{document}

\title{Inequivalent Hilbert space fragmentations from locally similar Hamiltonians}
\author{Somnath Maity}
\email{somnathmaity126@gmail.com}
\affiliation{Department of Physics, School of Basic Sciences, Indian Institute of Technology Bhubaneswar, Bhubaneswar 752050, India}
\author{Pramod Padmanabhan}
\email{pramod23phys@gmail.com}
\affiliation{Department of Physics, School of Basic Sciences, Indian Institute of Technology Bhubaneswar, Bhubaneswar 752050, India}
\author{Vladimir Korepin}
\email{vladimir.korepin@stonybrook.edu}
\affiliation{C. N. Yang Institute for Theoretical Physics, Stony Brook University, Stony Brook, New York 11794, USA}

\begin{abstract}
We construct two integrable, non-Hermitian spin-1 chains with open
boundaries by embedding a constant spin-\(\tfrac12\) Yang--Baxter
solution into the Hilbert space of two spin-1 sites in two different
ways and Baxterizing the resulting braid operators. A non-unitary
on-site similarity transformation relates the two braid operators,
their Baxterized \(R\)-matrices, and the corresponding double-row
transfer matrices. The two Hamiltonians are nilpotent and have the
same Jordan structure at every chain length, yet they fragment the
Hilbert space in inequivalent ways. In the standard spin inner
product, the similarity maps a conserved orthogonal projector of one
chain to an oblique projector that fails to commute with an adjoint
bond density of the other. Consequently, the similarity does not
intertwine the two bond \(*\)-algebras. In one chain, frozen two-site
barriers produce strong fragmentation. The largest orthogonal sector
has dimension \(F_{2L-1}\), a Fibonacci number, and occupies an
exponentially small fraction of the full Hilbert space. In the other
chain, fragmentation is weak, and this fraction is exactly \(2/9\)
for every \(L\geq 2\). With periodic boundaries the largest-sector
fraction tends to zero in the strongly fragmented chain and to one in
the weakly fragmented chain. These chains belong to a larger family of
integrable spin-1 models with Hilbert space fragmentation obtained
from braid-preserving embeddings of the non-invertible solution
\(R_{1,11}\) in Hietarinta's classification of constant \(4\times4\)
Yang--Baxter solutions.
\end{abstract}
\maketitle

\section{Introduction}\label{sec:introduction}

Local constraints can split the Hilbert space of a many-body system into
exponentially many dynamically disconnected sectors, even at fixed values
of the global conserved charges. This phenomenon is known as Hilbert space
fragmentation~\cite{Shattering,SalaDipole,HSFReview}. An early example
occurs in a model of diffusing dimers, where a conserved sequence of
labels, the irreducible string, distinguishes the sectors~\cite{DimerSectors}.
In quantum chains, charge and dipole conservation can restrict the allowed
moves~\cite{Shattering,SalaDipole}, and a pair-hopping Hamiltonian can
have sectors with different thermalization properties~\cite{PairKrylov}.
Experiments with ultracold atoms have probed fragmentation in tilted
Fermi--Hubbard chains~\cite{TiltedFermi} and observed it in a two-dimensional
tilted Bose--Hubbard system~\cite{TiltedBose}. Fragmentation is called
\emph{strong} if the largest fragment occupies a vanishing fraction of the
relevant symmetry sector as the system grows, and \emph{weak} if that
fraction remains bounded away from zero~\cite{SalaDipole,HSFReview}.

Fragmentation also occurs in integrable spin chains~\cite{FoldedSpinChain},
and in non-Hermitian chains it can help explain why the spectrum is
real~\cite{GhoshFragmentation}. Here we ask whether two spin chains
related by an on-site similarity transformation must fragment in the same
way. An orthogonal sector projector commutes with every bond density and
its adjoint~\cite{MM}. With the spin inner product held fixed, a
non-unitary similarity can carry such a projector to an oblique one.
We combine the Baxterization of a non-invertible Yang--Baxter
solution~\cite{MPK} with a construction that embeds operators on pairs
of sites~\cite{PK}. The resulting open spin-1 chains have identical
spectra at every length, yet one fragments strongly and the other weakly.

The rest of the paper is organized as follows.
Section~\ref{sec:two} constructs the two Hamiltonians from double-row
transfer matrices and gives the on-site similarity relating them.
Section~\ref{sec:fragmentation} identifies the orthogonal sectors of
each open chain and determines their dimensions. Section~\ref{sec:ring}
then examines the effect of closing the chain.
Section~\ref{sec:discussion} compares the two bond commutants and places
the models in a larger family of integrable chains.
Section~\ref{sec:outlook} outlines the open questions.
Appendix~\ref{app:F-count} derives the sector counts for the strongly
fragmented open chain. Appendix~\ref{app:ring} derives the periodic
sector counts and proves irreducibility of the active sectors.
Appendix~\ref{app:lifts} enumerates the computational-basis embeddings
that preserve the braid relation and classifies the resulting \(R\)-matrices.

\section{Two equivalent Hamiltonians}\label{sec:two}

Consider an open chain of \(L\geq2\) spin-1 sites. The local space
\(W=\mathbb C^3\) has orthonormal basis \(\ket0,\ket1,\ket2\), which we
take in order of decreasing \(S^z\) eigenvalue \(+1,0,-1\). We use the
standard inner product on \(W^{\otimes L}\) throughout. For nonzero
\(\alpha,\beta\in\mathbb C\), define the two bond densities
\begin{equation}
\begin{aligned}
h_{\mathrm{str}}&=\ket{01}\bigl(\alpha\bra{00}+\beta\bra{20}\bigr),\\
 h_{\mathrm{wk}}&=\ket{01}\bigl(\alpha\bra{00}+\beta\bra{02}\bigr).
\end{aligned}
\label{eq:densities}
\end{equation}
The subscripts \(\mathrm{str}\) and \(\mathrm{wk}\) anticipate the strong
and weak Hilbert space fragmentation, respectively, that we establish in
Section~\ref{sec:fragmentation} using the standard spin inner product.
Both densities send \(\ket{00}\) to \(\alpha\ket{01}\). The second input is
\(\ket{20}\) for \(h_{\mathrm{str}}\) and \(\ket{02}\) for \(h_{\mathrm{wk}}\), in
either case with amplitude \(\beta\). All other computational-basis inputs
are annihilated. The two non-Hermitian Hamiltonians are
\begin{equation}
\begin{aligned}
H_\sigma&=\sum_{j=1}^{L-1}h_{\sigma,j},\qquad \sigma\in\{\mathrm{str},\mathrm{wk}\},\\
 h_{\sigma,j}&=\id^{\otimes(j-1)}\otimes h_\sigma\otimes\id^{\otimes(L-j-1)}.
\end{aligned}
\label{eq:H}
\end{equation}
There is no bond joining sites \(L\) and \(1\). The orientation in
\eqref{eq:densities} fixes which spin is the first argument of each bond.

\subsection{Yang--Baxter integrability}

Both densities descend from the rank-one class \(R_{1,11}\) in the
classification of constant \(4\times4\) Yang--Baxter solutions~\cite{Hietarinta}. In the ordered two-spin-\(\tfrac12\) basis
\((00,01,10,11)\), the representative and its Baxterized local Hamiltonian
density have the form \cite[Sec.~4.8, Eqs.~(4.44)--(4.48)]{MPK}
\begin{equation}
 h_{1/2}=\ket{00}\bigl(\alpha\bra{01}+\beta\bra{10}\bigr),\qquad h_{1/2}^2=0,
 \label{eq:seed}
\end{equation}
up to the overall Hamiltonian normalization. To specify the spin-1
realizations, let \(J_{\sigma}:(\mathbb C^2)^{\otimes2}\to W^{\otimes2}\)
be the bond isometry whose images of those four basis kets are
\begin{align*}
 J_{\mathrm{str}}:&\quad(\ket{01},\ket{00},\ket{20},\ket{22}),\\
 J_{\mathrm{wk}}:&\quad(\ket{01},\ket{00},\ket{02},\ket{22}).
\end{align*}
Then \(J_\sigma h_{1/2}J_\sigma^\dagger=h_\sigma\), giving
\eqref{eq:densities} by the bond-wise construction of Ref.~\cite{PK}.
Since two adjacent bond embeddings share a spin, the three-site
Yang--Baxter equation must be checked in \(W^{\otimes3}\).

For this check write \(\bra{s}=\alpha\bra0+\beta\bra2\) and
\(h_\sigma=X_\sigma\otimes Y_\sigma\). The one-site factors are
\begin{equation}
\begin{array}{c|cc}
 \sigma &X_\sigma&Y_\sigma\\ \hline
 \mathrm{str}&\ket0\bra{s}&\ket1\bra0\\[2pt]
 \mathrm{wk}&\ket0\bra0&\ket1\bra{s}
\end{array}
\label{eq:factors}
\end{equation}
In each row \(Y_\sigma^2=X_\sigma Y_\sigma=0\). The first identity gives
\(h_{\sigma,j}^2=0\). The second gives
\(h_{\sigma,j+1}h_{\sigma,j}=0\), since the factor on the common site is
\(X_\sigma Y_\sigma\). Thus, writing \(A=h_{\sigma,1}\) and
\(B=h_{\sigma,2}\), we have \(A^2=B^2=0\) and \(ABA=BAB=0\).
The linear Baxterization \cite{MPK} becomes
\begin{equation}
 \check R_\sigma(u)=\id_9+u h_\sigma,\qquad
 R_{\sigma,12}(u)=P_{12}\check R_{\sigma,12}(u),
 \label{eq:R}
\end{equation}
where \(P_{12}\ket{ij}=\ket{ji}\). Indeed, the braided spectral equation is
\begin{equation}
\begin{aligned}
&(\id+uA)(\id+(u+v)B)(\id+vA)\\
 &\qquad=(\id+vB)(\id+(u+v)A)(\id+uB).
\end{aligned}
\label{eq:braid}
\end{equation}
The linear terms agree. The difference at quadratic order is
\(uv(A^2-B^2)\), and at cubic order it is
\(uv(u+v)(ABA-BAB)\), both zero. Consequently \(R_\sigma(u)\) solves the
ordinary difference-form Yang--Baxter equation. It is regular at \(u=0\)
and satisfies the inversion identity
\begin{equation}
\begin{aligned}
R_{\sigma,12}(0)&=P_{12},\qquad R_{\sigma,12}(u)R_{\sigma,21}(-u)=\id_9,\\
 R_{\sigma,21}(u)&=P_{12}R_{\sigma,12}(u)P_{12}.
\end{aligned}
\label{eq:inversion}
\end{equation}

In the quantum inverse scattering method (QISM)~\cite{SF78,STF79},
the single-row monodromy acts on an auxiliary spin \(a\) and the chain.
Its untwisted auxiliary trace gives the transfer matrix~\cite{IK,KBI},
\begin{equation}
 T_{\sigma,a}(u)=R_{\sigma,aL}(u)\cdots R_{\sigma,a1}(u),\qquad
 t_\sigma(u)=\tr_a T_{\sigma,a}(u).
 \label{eq:periodic-transfer}
\end{equation}
The Yang--Baxter equation gives the exchange relation for two such
monodromies. Tracing over both auxiliary spaces yields
\([t_\sigma(u),t_\sigma(v)]=0\), as in the algebraic Bethe ansatz
\cite{SlavnovReview,SlavnovLectures,SlavnovBook}. The trace in
\eqref{eq:periodic-transfer} closes the row. At \(u=0\) each \(R\)-matrix
is a swap, so \(t_\sigma(0)\) shifts the spin configuration by one site,
\(\ket{s_1\cdots s_L}\mapsto\ket{s_Ls_1\cdots s_{L-1}}\).
Differentiating one factor at a time and cancelling this shift places
\(\check R_\sigma'(0)=h_\sigma\) on each oriented bond around the ring.
Hence the single-row transfer matrix generates the \emph{periodic} Hamiltonian
\begin{equation}
 H_\sigma^{\rm per}=t_\sigma(0)^{-1}t_\sigma'(0)
 =\sum_{j=1}^{L-1} h_{\sigma,j}+(h_\sigma)_{L,1}
 =H_\sigma+(h_\sigma)_{L,1}.
 \label{eq:periodic-H}
\end{equation}
The closing bond acts on site \(L\) as its first argument and site \(1\) as
its second. To generate the open Hamiltonian \eqref{eq:H} from a commuting
transfer matrix we introduce reflecting boundaries.

\subsection{Reflecting boundaries and the open-chain transfer matrix}

The reflection equation requires the two possible orders of scattering
and reflection for two particles on a half-line to give the same
amplitude~\cite{Cherednik}. This condition yields a commuting family
of double-row transfer matrices for an open chain~\cite{Sklyanin}.
We apply this construction to each \(R_\sigma\)
and suppress \(\sigma\) in this subsection. Let \(a,b\) denote auxiliary
copies of \(W\). At the left end the reflection equation is
\begin{equation}
\begin{aligned}
&R_{ab}(u-v)K^-_a(u)R_{ba}(u+v)K^-_b(v)\\
 &\qquad=K^-_b(v)R_{ab}(u+v)K^-_a(u)R_{ba}(u-v).
\end{aligned}
\label{eq:RE}
\end{equation}
For each of our two \(R\)-matrices, take \(K^-(u)=\id_3\). By \eqref{eq:R},
\(R_{ab}(x)R_{ba}(y)=\id+(x+y)h_{ba}\), so both sides of
\eqref{eq:RE} are \(\id+2u h_{ba}\). At the right end choose
\(K^+(u)=\id_3\) as the matrix in the auxiliary trace. We verify this
choice directly below. The usual dual-reflection argument uses
\((R_{ab}^{t_a})^{-1}\) at generic spectral parameter
\cite[Theorem~2.4]{Vlaar}. Here \(P_{ab}^{t_a}\) has rank one, and
\(P_{ab}h_{ab}\) is a product of rank-one one-site operators in both
models. Thus \(R_{ab}^{t_a}(u)\) has rank at most two on the
nine-dimensional auxiliary pair for every \(u\).

Using the monodromy in \eqref{eq:periodic-transfer}, define the dressed
boundary operator and normalized double-row transfer matrix \cite{Sklyanin} by
\begin{equation}
\begin{aligned}
U_a(u)&=T_a(u)K^-_a(u)T_a^{-1}(-u),\\
 \tau_L(u)&=\tfrac13\tr_a\!\left[K^+_a(u)U_a(u)\right].
\end{aligned}
\label{eq:double}
\end{equation}
The inverse row in \eqref{eq:double} is
\(R_{1a}(u)\cdots R_{La}(u)\) by \eqref{eq:inversion}. To evaluate the
trace, write \(\check R_j(u)=\check R_{j,j+1}(u)=\id+u h_j\) and define the operator on the physical
chain
\begin{equation}
 D_L(u)=\check R_{L-1}(u)\cdots\check R_1(u)\,\check R_1(u)\cdots\check R_{L-1}(u).
 \label{eq:palindrome}
\end{equation}
Moving the permutations through the two rows gives
\begin{equation}
 U_a(u)=\bigl(\id+uX_LY_a\bigr)\,D_L(u)\,
         \bigl(\id+uX_LY_a\bigr).
 \label{eq:fold}
\end{equation}
One way to see the ordering is to add the sites successively.
Adding site \(j\) surrounds the previous double row by \(R_{aj}\) and
\(R_{ja}\). Using \(R_{aj}=\check R_{ja}P_{aj}\) and
\(R_{ja}=P_{aj}\check R_{ja}\), the two permutations replace the old
auxiliary label by \(j\). The new outer factors are both \(\check R_{ja}\).
Starting with \(K^-=\id\) yields \eqref{eq:palindrome}--\eqref{eq:fold}.
Since \(D_L\) acts trivially in \(a\), the term quadratic in the outer
factors vanishes by \(Y_a^2=0\). The linear terms have zero auxiliary
trace because \(\tr Y=0\). Therefore \(\tau_L(u)=D_L(u)\).

To establish commutativity, we use the dressing identity~\cite{Sklyanin}.
It promotes the numerical solution \(K^-\) to an operator-valued
solution \(U_a\) of the reflection equation. The entries of \(U_a\)
then satisfy the quadratic relations of the reflection
algebra~\cite{KulishSklyanin}. Moving the monodromies through \(R_{ab}\) by
the Yang--Baxter equation gives the operator identity
\begin{equation}
\begin{aligned}
&R_{ab}(u-v)U_a(u)R_{ba}(u+v)U_b(v)\\
 &\qquad=U_b(v)R_{ab}(u+v)U_a(u)R_{ba}(u-v).
\end{aligned}
\label{eq:dressedRE}
\end{equation}
For both factorizations in \eqref{eq:factors},
\(Y\ket1=0\). Thus \eqref{eq:fold} implies, for every physical
ket \(\ket\psi\),
\[
 U_a(u)(\ket\psi\otimes\ket1_a)=
       D_L(u)\ket\psi\otimes\ket1_a .
\]
Moreover \(R_{ab}(w)\) and \(R_{ba}(w)\) fix \(\ket{1_a1_b}\) at every
\(w\). Restricting \eqref{eq:dressedRE} to this auxiliary line
gives \(D_L(u)D_L(v)=D_L(v)D_L(u)\). Thus the restriction to
\(\ket{1_a1_b}\) proves commutativity of the ordinary traces with
\(K^+=\id_3\) for both chains.

Restoring the model index, every physical bond in \eqref{eq:palindrome}
occurs once in each row. Since \(\check R_{\sigma,j}(0)=\id\) and
\(\check R_{\sigma,j}'(0)=h_{\sigma,j}\), it follows that
\begin{equation}
\begin{aligned}
\bigl[\tau_{\sigma,L}(u),\tau_{\sigma,L}(v)\bigr]&=0,\qquad \tau_{\sigma,L}(0)=\id,\\
 \left.\frac12\frac{d}{du}\log\tau_{\sigma,L}(u)\right|_{u=0}
 &=\sum_{j=1}^{L-1}h_{\sigma,j}=H_\sigma .
\end{aligned}
\label{eq:logder}
\end{equation}
The auxiliary trace removes the last auxiliary bond exactly. Thus the
logarithmic derivative generates the open Hamiltonians
\eqref{eq:H} with zero one-site boundary fields.

\subsection{A homogeneous local similarity}\label{sec:local-similarity}

The two regular \(R\)-matrices belong to the same class under a
homogeneous change of one-site basis \cite{Hietarinta}. Set
\begin{equation}
\begin{aligned}
G&=\begin{pmatrix}1&0&t\\0&1&0\\0&0&-1\end{pmatrix},\qquad t=\frac{\beta}{\alpha},\\
 G^{-1}&=G,\qquad S_L=G^{\otimes L}.
\end{aligned}
\label{eq:G}
\end{equation}
Recall that \(\ket{s}=\overline{\alpha}\ket0+\overline{\beta}\ket2\),
so \(\bra{s}=\alpha\bra0+\beta\bra2\).
The map fixes \(\ket0,\ket1\), while
\(\bra0G=\bra{s}/\alpha\) and \(\bra{s}G=\alpha\bra0\).
The factors of \(\alpha\) cancel between the two sites of \(h_{\mathrm{str}}\).
Since \(G\otimes G\) also commutes with the swap, one obtains
\begin{align}
 h_{\mathrm{wk}}&=(G\otimes G)h_{\mathrm{str}}(G\otimes G)^{-1},\nonumber\\
 R_{\mathrm{wk}}(u)&=(G\otimes G)R_{\mathrm{str}}(u)(G\otimes G)^{-1}.
 \label{eq:gaugeR}
\end{align}
Both identity boundary matrices are fixed by \(G\). Conjugating the full
double row and taking the auxiliary trace cancels the auxiliary factors
\(G_a\) and \(G_a^{-1}\), leaving
\begin{equation}
 \tau_{\mathrm{wk},L}(u)=S_L\tau_{\mathrm{str},L}(u)S_L^{-1},\qquad
 H_{\mathrm{wk}}=S_L H_{\mathrm{str}}S_L^{-1}.
 \label{eq:gaugeH}
\end{equation}
Thus all transfer-matrix charges are related by the same similarity, and
the two Hamiltonians have identical Jordan form at every length.

\section{Two inequivalent Hilbert space fragmentations}\label{sec:fragmentation}

We will now show that both \(H_{\mathrm{str}}\) and \(H_{\mathrm{wk}}\) fragment the
Hilbert space. Given the on-site similarity transformation established
in Section~\ref{sec:local-similarity}, one might expect the two Hamiltonians
to exhibit similar fragmentation mechanisms. This raises the question:
\emph{Does the local similarity also carry the orthogonal fragmentation sectors
of \(H_{\mathrm{str}}\) into those of \(H_{\mathrm{wk}}\)?}

Before examining the two Hamiltonians in turn, we specify what defines
a fragmentation sector for either chain. We use the same spin inner
product on \(\mathcal H_L=W^{\otimes L}\) for both. For each
\(\sigma\in\{\mathrm{str},\mathrm{wk}\}\), the local densities and their adjoints
generate a unital bond \(*\)-algebra \(\mathfrak B_{\sigma,L}\).
Its commutant \(\mathfrak C_{\sigma,L}\) consists of the operators that
commute with all those generators~\cite{MM}:
\begin{equation}
\begin{aligned}
\mathfrak B_{\sigma,L}
 &=\operatorname{alg}_{\mathbb C}\{\id,h_{\sigma,j},h_{\sigma,j}^{\dagger}:1\le j<L\},\\
 \mathfrak C_{\sigma,L}
 &=\mathfrak B_{\sigma,L}'=\bigl\{Q\in\operatorname{End}(\mathcal H_L):\\[-2pt]
 &\qquad [Q,h_{\sigma,j}]=[Q,h_{\sigma,j}^{\dagger}]=0\ \text{for every }j\bigr\}.
\end{aligned}
\label{eq:sector-algebras}
\end{equation}
Here \(\operatorname{alg}_{\mathbb C}\) denotes the complex linear span of
all finite products of the displayed generators.
An orthogonal projector \(Q=Q^\dagger=Q^2\) belongs to
\(\mathfrak C_{\sigma,L}\) precisely when every bond preserves both
its range and its orthogonal complement. We take the irreducible sectors
to be the minimal nonzero subspaces with this property. A \emph{frozen}
state is annihilated by every \(h_{\sigma,j}\) and \(h_{\sigma,j}^\dagger\).
Its span is a one-dimensional sector of energy zero. This common-kernel
condition is stronger than \(H_\sigma\ket\psi=0\).

If \(Q\) projects onto a sector
of \(H_{\mathrm{str}}\), its image \(S_L Q S_L^{-1}\) is still idempotent and
commutes with every bond of \(H_{\mathrm{wk}}\). Since the local map \(G\) is
nonunitary, \(S_L\) can carry the range and kernel of \(Q\) into
nonorthogonal subspaces, making the transformed projector oblique.
The two chains can therefore admit different orthogonal sector
decompositions despite their local similarity. In the next two
subsections we determine the sectors of \(H_{\mathrm{str}}\) and \(H_{\mathrm{wk}}\)
in turn and identify the mechanism responsible for fragmentation in each.

\subsection{Fragmentation of \texorpdfstring{\(H_{\mathrm{str}}\)}{H\textunderscore str}}\label{sec:F-sectors}

In this subsection write \(h_j=h_{\mathrm{str},j}\). On a bond,
\(h_{\mathrm{str}}\) has two nonzero computational-basis inputs, and
\(h_{\mathrm{str}}^\dagger\) has one. Their actions follow from \eqref{eq:densities}:
\begin{equation}
\begin{aligned}
h_{\mathrm{str}}\ket{00}&=\alpha\ket{01},\\
h_{\mathrm{str}}\ket{20}&=\beta\ket{01},\\
h_{\mathrm{str}}^\dagger\ket{01}&=\overline\alpha\ket{00}+\overline\beta\ket{20}.
\end{aligned}
\label{eq:F-moves}
\end{equation}
Each operator annihilates all its other basis inputs. In particular the
adjoint produces a \emph{superposition} with a definite relative amplitude.
To see which linear combination participates, introduce
\begin{equation}
\begin{aligned}
N&=\sqrt{|\alpha|^2+|\beta|^2},\\
 \ket v&=\frac{\overline\alpha\ket0+\overline\beta\ket2}{N},\qquad
 \ket z=\frac{\beta\ket0-\alpha\ket2}{N}.
\end{aligned}
\label{eq:F-vz}
\end{equation}
The kets \(\ket v=\ket s/N\) and \(\ket z\) are an orthonormal basis of
\(V=\operatorname{span}\{\ket0,\ket2\}\). To see which states a single bond connects, consider the two-site
chain. In this basis the bond operator and its adjoint connect only
\(\ket{v0}\) and \(\ket{01}\), since
\[
 h_{\mathrm{str}}=N\ket{01}\bra{v0},\qquad
 h_{\mathrm{str}}^\dagger=N\ket{v0}\bra{01}.
\]
On a longer chain, adjacent bonds act on a common spin. A ket killed by
one bond can therefore still be changed by its neighbor.

To idenitfy the orthogonal sectors of the Hilbert space we begin by showing that a fixed \(21\) splits the chain into dynamically disconnected segments.
Suppose sites \(j,j+1\) are in \(\ket{21}\). The bond \(h_j\) and its
adjoint both kill this pair by \eqref{eq:F-moves}. On the bond directly
to the left, the \emph{right} spin is \(2\), whereas both moves require
that spin to be \(0\) or \(1\). On the bond directly to the right, the
\emph{left} spin is \(1\), whereas both moves require that spin to lie
in \(V\). Thus all bonds touching this pair vanish on
\(\ket{\psi_{\rm left}}\otimes\ket{21}\otimes\ket{\psi_{\rm right}}\),
whatever the states on its two sides. The other bonds act entirely on
one side. This is the sense in which \(21\) is a barrier.

Let \(B_j=(\ket{21}\bra{21})_{j,j+1}\), with identities elsewhere.
For \(k=j-1,j,j+1\) within the chain, the preceding argument gives
\(h_kB_j=h_k^\dagger B_j=0\). Taking adjoints gives
\(B_jh_k=B_jh_k^\dagger=0\). Disjoint supports commute, so
\begin{equation}
 B_j\in\mathfrak C_{\mathrm{str},L}\qquad (1\le j<L).
 \label{eq:F-barrier}
\end{equation}
This also gives a concrete answer to the question opening Section~\ref{sec:fragmentation}.
Under the similarity \eqref{eq:G} the two-site factor of \(B_j\) becomes
\((\ket2-t\ket0)\bra2\otimes\ket1\bra1\), an oblique projector for \(t\ne0\).

The \(B_j\) are diagonal in the word basis and commute with one another.
An \emph{exact} pattern specifies both where a barrier occurs and where
it is absent. Its projector is
\begin{equation}
 \Pi_S=\prod_{j\in S}B_j\prod_{j\notin S}(\id-B_j),
 \qquad S\subseteq\{1,\ldots,L-1\}.
 \label{eq:F-pattern}
\end{equation}
No two members of \(S\) can be consecutive in a nonzero pattern, since
their common spin would have to be both \(1\) and \(2\). Between
successive fixed pairs, and between a fixed pair and an open end, the
remaining words contain no \(21\). Each such interval can be analyzed
independently. For a concrete example of this cut, fix \(21\) on sites \(3,4\) of a
six-site chain. This leaves the bonds \((1,2)\) and \((5,6)\) acting on separate two-site
intervals. If this is the only barrier, \(21\) is excluded in each interval.

Next, we remove the frozen ends of one interval.
Consider an interval of \(n\) sites containing no \(21\). First look at
its leftmost spin. If it is \(1\), the first bond kills the state in
both directions, and all remaining bonds leave this spin alone. The
projector onto this choice therefore commutes with the restricted bond
algebra. Removing that spin exposes another left end with the same
rule. Repetition gives a prefix \(\ket1^{\otimes a}\). Once we take the
orthogonal alternative, the first surviving spin lies in \(V\).

At the right end a spin \(2\) is likewise untouched, since both bond
moves act only between \(0\) and \(1\) on their right leg. Repeating
this choice gives a suffix \(\ket2^{\otimes c}\). If an end remains,
the last surviving spin now lies in
\(\operatorname{span}\{\ket0,\ket1\}\). If \(a+c=n\), the prefix and
suffix fill the entire interval, giving the frozen configuration \(1^a2^c\).

There is one more way for the left end to freeze. Suppose at least two
sites remain, and inspect the second spin. If it is \(1\), the first
spin, already in \(V\), must be \(0\) because the word \(21\) is
excluded. If the second spin lies in \(V\), we can resolve the first
spin along \(\ket v,\ket z\). Hence the allowed first-pair space at this
stage splits orthogonally as
\begin{equation}
 \bigl(\mathbb C\ket z\otimes V\bigr)
 \ \oplus\
 \underbrace{\bigl[\operatorname{span}\{\ket{01}\}
                 \oplus(\mathbb C\ket v\otimes V)\bigr]}_{\mathcal E}.
 \label{eq:F-left-branches}
\end{equation}
These first-pair conditions are understood together with the restrictions
on the rest of the interval. To see why the first summand freezes the
left spin, take any \(\ket w\in V\). The operator \(h_1\) kills
\(\ket{zw}\) because \(\langle v|z\rangle=0\), and \(h_1^\dagger\)
kills it because \(\langle1|w\rangle=0\). The next bond and its adjoint
preserve \(V\) on their left leg. Consequently the second spin stays in
\(V\), and the first spin stays in \(\ket z\). We may remove this first
spin and repeat the split at its neighbor. The two-site example \(h_{\mathrm{str}}^\dagger\ket{z1}=\beta\ket{v0}\)
shows why the neighbor must lie in \(V\) for this removal to be valid. Successive choices of the
\(\mathbb C\ket z\otimes V\) branch give a frozen string
\(\ket z^{\otimes b}\) at the left edge of the remaining interval.
If one spin finally remains, it belongs to
\(V\cap\operatorname{span}\{\ket0,\ket1\}=\mathbb C\ket0\).

The second summand \(\mathcal E\) in \eqref{eq:F-left-branches} is the
orthogonal complement of the frozen branch within the allowed first-pair
space. It keeps the transition \(\ket{v0}\leftrightarrow\ket{01}\)
at the left edge. The resulting interval sectors take the following forms.
We use a string of kets to denote their tensor product and omit a factor
when its exponent is zero:
\begin{align}
 &\ket1^{\otimes a}\ket2^{\otimes c},
       &&a+c=n, \label{eq:F-frozen-ends}\\
 &\ket1^{\otimes a}\ket z^{\otimes b}\ket0\ket2^{\otimes c},
       &&a+b+1+c=n, \label{eq:F-frozen}\\[2pt]
 &\ket1^{\otimes a}\ket z^{\otimes b}\otimes\mathcal A_m
        \otimes\ket2^{\otimes c},
       &&a+b+m+c=n,\quad m\ge2. \label{eq:F-active}
\end{align}
Here \(a,b,c\) are nonnegative. In \eqref{eq:F-active}, the \(m\)
consecutive sites left after removing the frozen end spins form the
\emph{active segment}. Its allowed states span \(\mathcal A_m\).
Each ket in \eqref{eq:F-frozen-ends} or
\eqref{eq:F-frozen} spans a
frozen line. These kets give one orthonormal choice within the common
zero space, where a change of orthonormal basis is always possible.
For an empty interval the first line contributes the empty tensor product.

Now we identify the active space \(\mathcal A_m\).
Its first pair lies in \(\mathcal E\), its last spin is \(0\) or \(1\),
and it contains no barrier. These conditions give an orthonormal basis
without diagonalizing a Hamiltonian. Choose a computational suffix
\(t_2\cdots t_m\) with no occurrence of \(21\) and \(t_m\in\{0,1\}\).
The second spin then fixes the first ket:
\begin{equation}
\begin{aligned}
&\ket{\chi(t_2)}\otimes\ket{t_2\cdots t_m},\\
 &\ket{\chi(1)}=\ket0,\qquad \ket{\chi(0)}=\ket{\chi(2)}=\ket v .
\end{aligned}
\label{eq:F-core-basis}
\end{equation}
Distinct suffixes are orthogonal even though \(\ket0\) and \(\ket v\)
have nonzero overlap.

To illustrate the basis construction in \eqref{eq:F-core-basis}, consider
active segments of lengths two and three. For \(m=2\) the suffix consists
of a single \(0\) or \(1\), giving
\(\mathcal A_2=\operatorname{span}\{\ket{v0},\ket{01}\}\).
For \(m=3\) the allowed suffixes are \(00,01,10,11,20\), so
\[
 \mathcal A_3=\operatorname{span}
 \{\ket{v00},\ket{v01},\ket{010},\ket{011},\ket{v20}\}.
\]
To see how neighboring bonds act within this three-site example, start
from \(\ket{011}\). The bond on sites \(2,3\) kills it in both directions,
but \(h_1^\dagger\ket{011}=N\ket{v01}\). In turn
\(h_2^\dagger\ket{v01}=\overline\alpha\ket{v00}+\overline\beta\ket{v20}\),
and \(h_1\ket{v00}=N\ket{010}\). Thus a bond that is inactive on one
ket can participate after its neighbor acts.

To illustrate how a frozen prefix fits together with an active segment,
take \(a=b=1\), \(c=0\) and \(m=2\) in \eqref{eq:F-active}. This gives
the four-site sector \(\ket{1z}\otimes\mathcal A_2\), with sites 3 and 4 active. A basis of the sector is
\(\ket{1zv0},\ket{1z01}\). The first two bonds vanish in both directions,
and the third bond and its adjoint connect these two kets.

Now we show that the active space is a single sector.
The end restrictions ensure that every \(h_j\) and \(h_j^\dagger\) preserves
\(\mathcal A_m\). Could \(\mathcal A_m\) itself be an orthogonal sum of two
smaller sectors? If so, the orthogonal projector \(Q\) onto either summand
would commute with every bond and its adjoint on \(\mathcal A_m\). We show
that such a projector must be zero or the identity on \(\mathcal A_m\).

\emph{A bond product singles out one ket.}
Choose \(\ket{p_m}=\ket0\ket1^{\otimes(m-1)}\), a basis ket in
\eqref{eq:F-core-basis}. Each bond has the factorized form
\(h_j=N(\ket0\bra v)_j(\ket1\bra0)_{j+1}\).
In the ordered product \(h_1h_2\cdots h_{m-1}\), the two factors on each
interior site multiply to \((\ket1\bra0)(\ket0\bra v)=\ket1\bra v\).
The first and last sites retain their original factors, giving
\begin{equation}
 T_m=h_1h_2\cdots h_{m-1}
     =N^{m-1}\ket{p_m}\bigl(\bra v^{\otimes(m-1)}\otimes\bra0\bigr).
 \label{eq:F-rank-one}
\end{equation}
Since \(\ket v\) is normalized, the bond \(*\)-algebra therefore contains
the rank-one projector \(P_m=T_mT_m^\dagger/N^{2(m-1)}
=\ket{p_m}\bra{p_m}\).
From \(QP_m=P_mQ\) we get
\(Q\ket{p_m}=P_mQ\ket{p_m}=\lambda\ket{p_m}\).
The identity \(Q^2=Q\) then forces \(\lambda=0\) or \(1\).
For every bond-algebra element \(X\), commutation gives
\[
 QX\ket{p_m}=XQ\ket{p_m}=\lambda X\ket{p_m}.
\]
Thus \(Q\) acts with the \emph{same} eigenvalue on the entire linear span
of the vectors \(X\ket{p_m}\). It remains to prove that this span is all of
\(\mathcal A_m\). A ket with this spanning property is called cyclic.

\emph{How the bond operators and their adjoints generate \(\mathcal A_m\) from \(\ket{p_m}\).}
We first work out how the operators generate the basis on the two
shortest active segments. The two-site example gives the base case of
the induction. \(\ket{p_2}=\ket{01}\) and
\(h_1^\dagger\ket{01}=N\ket{v0}\) give the two basis kets.
The three-site example shows why we take the linear span. Starting with
\(\ket{p_3}=\ket{011}\), the moves above give \(\ket{v01}\) and then
\(u=\overline\alpha\ket{v00}+\overline\beta\ket{v20}\).
Since \(h_1\) kills the second term,
\(h_1u=N\overline\alpha\ket{010}\) supplies \(\ket{010}\), and
\(h_1^\dagger\ket{010}=N\ket{v00}\) supplies \(\ket{v00}\).
Subtracting \(\overline\alpha\ket{v00}\) from \(u\) isolates
\(\overline\beta\ket{v20}\). This accounts for all five basis kets of
\(\mathcal A_3\), using \(\alpha\beta\ne0\).

For the induction, assume \(\ket{p_{m-1}}\) is cyclic on \(m-1\) sites,
with \(m\ge3\). Sort the basis \eqref{eq:F-core-basis} by its last two spins.
\begin{enumerate}
\item \emph{Kets ending in \(1\).}
The penultimate spin is \(0\) or \(1\), since \(21\) is forbidden.
Deleting the last spin leaves precisely a basis ket of \(\mathcal A_{m-1}\).
Starting from \(\ket{p_m}=\ket{p_{m-1}}\ket1\), the bonds on the
first \(m-1\) sites therefore generate all of
\(\mathcal A_{m-1}\otimes\ket1\), with the final spin untouched.

\item \emph{Kets ending in \(00\) or \(10\).}
These form the basis of \(\mathcal A_{m-1}\otimes\ket0\). To use the
same induction, we first need \(\ket{p_{m-1}}\ket0\).
Equation~\eqref{eq:F-rank-one} gives the two successive actions
\[
\begin{aligned}
 T_m^\dagger\ket{p_m}
   &=N^{m-1}\ket v^{\otimes(m-1)}\ket0,\\
 (T_{m-1}\otimes\id)\bigl(\ket v^{\otimes(m-1)}\ket0\bigr)
   &=N^{m-2}\langle0|v\rangle\,\ket{p_{m-1}}\ket0 .
\end{aligned}
\]
The overlap is \(\langle0|v\rangle=\overline\alpha/N\ne0\).
Acting on the first \(m-1\) sites now generates this entire group.

\item \emph{Kets ending in \(20\).}
For each allowed \(\ket{w20}\), the kets \(\ket{w01}\) and
\(\ket{w00}\) occur in the preceding two groups. Replacing the
penultimate \(2\) by \(0\) preserves the suffix constraint. For \(m=3\)
it also preserves the first ket because \(\chi(2)=\chi(0)=v\).
The last bond then isolates the desired vector by linear combination,
\[
 \ket{w20}=\frac{h_{m-1}^\dagger\ket{w01}
                  -\overline\alpha\ket{w00}}{\overline\beta}.
\]
\end{enumerate}
Every active basis ket ends in \(0\) or \(1\), so the three groups exhaust
\eqref{eq:F-core-basis}. This proves cyclicity for every \(m\ge2\).
Consequently \(Q=\lambda\id\) on \(\mathcal A_m\), and the active space
is a single irreducible sector of the bond \(*\)-algebra.

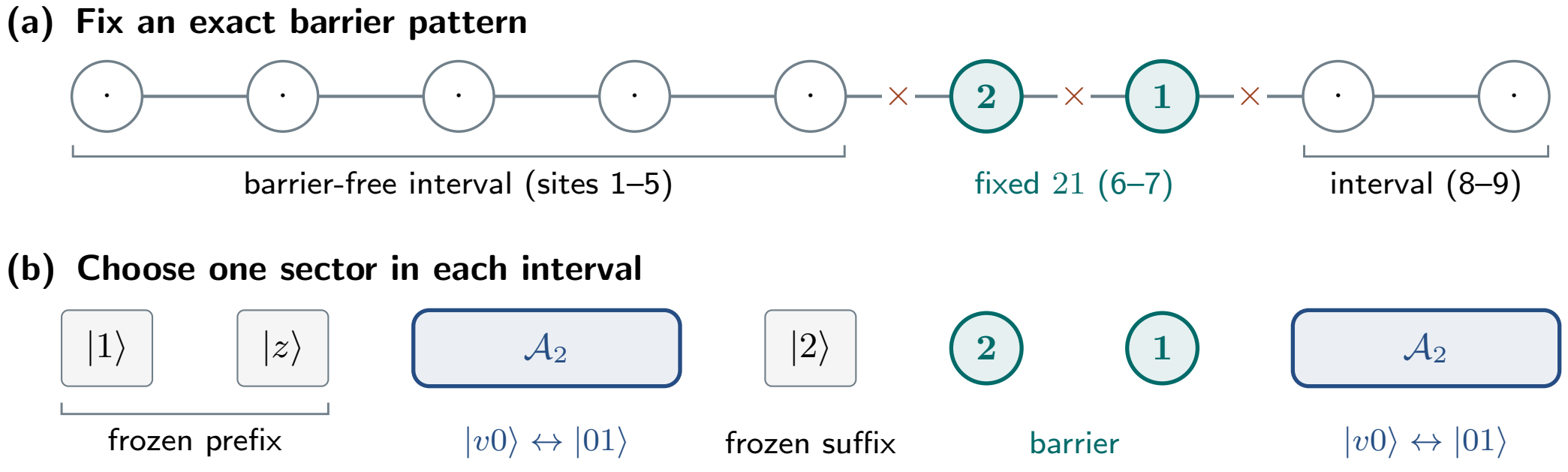
\begin{figure*}[t]
\centering
\begin{tikzpicture}[
  x=1cm,y=1cm,
  panel/.style={anchor=west,font=\small\sffamily\bfseries,inner sep=0pt},
  labeltext/.style={font=\footnotesize\sffamily,inner sep=1pt},
  site/.style={circle,draw=HFgray,line width=.65pt,fill=white,
               minimum size=6.2mm,inner sep=0pt,font=\small},
  barrier/.style={site,draw=HFbar,fill=HFbar!10,text=HFbar,line width=1pt},
  fixed/.style={draw=HFgray,fill=black!4,rounded corners=.6mm,
                minimum width=8mm,minimum height=6.7mm,inner sep=1pt},
  active/.style={draw=HFact,fill=HFact!9,text=HFact,rounded corners=1.4mm,
                 minimum width=23.5mm,minimum height=6.7mm,line width=1pt},
  crossed/.style={fill=white,inner sep=.5pt,text=HFcut,font=\small},
  group/.style={draw=HFgray,line width=.55pt}]
  \node[panel] at (0,0) {(a)\enspace Fix an exact barrier pattern};
  \draw[HFgray,line width=.8pt] (.9,-.64)--(13.3,-.64);
  \foreach \x in {.9,2.45,4,5.55,7.1,11.75,13.3}
    \node[site] at (\x,-.64) {$\cdot$};
  \node[barrier] at (8.65,-.64) {$\mathbf 2$};
  \node[barrier] at (10.2,-.64) {$\mathbf 1$};
  \foreach \x in {7.875,9.425,10.975}
    \node[crossed] at (\x,-.64) {$\times$};
  \draw[group] (.6,-1.07) |- (7.4,-1.17) -- (7.4,-1.07);
  \draw[group] (11.45,-1.07) |- (13.6,-1.17) -- (13.6,-1.07);
  \node[labeltext] at (4,-1.43) {barrier-free interval (sites 1--5)};
  \node[labeltext,text=HFbar] at (9.425,-1.43) {fixed $21$ (6--7)};
  \node[labeltext] at (12.525,-1.43) {interval (8--9)};

  \node[panel] at (0,-2.18) {(b)\enspace Choose one sector in each interval};
  \foreach \x/\s in {.9/1,2.45/z,7.1/2}
    \node[fixed] at (\x,-2.86) {$\ket{\s}$};
  \node[active] at (4.775,-2.86) {$\mathcal A_2$};
  \node[barrier] at (8.65,-2.86) {$\mathbf 2$};
  \node[barrier] at (10.2,-2.86) {$\mathbf 1$};
  \node[active] at (12.525,-2.86) {$\mathcal A_2$};
  \draw[group] (.5,-3.34) |- (2.85,-3.44) -- (2.85,-3.34);
  \node[labeltext] at (1.675,-3.69) {frozen prefix};
  \node[labeltext,text=HFact] at (4.775,-3.69) {$\ket{v0}\leftrightarrow\ket{01}$};
  \node[labeltext] at (7.1,-3.69) {frozen suffix};
  \node[labeltext,text=HFbar] at (9.425,-3.69) {barrier};
  \node[labeltext,text=HFact] at (12.525,-3.69) {$\ket{v0}\leftrightarrow\ket{01}$};
\end{tikzpicture}
\caption{Some orthogonal sectors on a length 9 open chain.}
\label{fig:F-sectors}
\end{figure*}

To specify a sector of the full chain, fix an exact barrier pattern, choose
one line or active sector from \eqref{eq:F-frozen-ends}--\eqref{eq:F-active}
in each intervening interval, and take their ordered tensor product with
\(\ket{21}\) at the fixed bonds. Different barrier patterns are orthogonal
by \eqref{eq:F-pattern}. Within one pattern, distinct choices separate at
an orthogonal end split, and the bonds act independently on the active
segments. Figure~\ref{fig:F-sectors} illustrates this construction on a nine-site
chain, from the choice of barrier to the choice of a sector in each interval. For each pattern \(S\), the orthogonal sum over the interval choices
gives \(\operatorname{ran}\Pi_S\). Summing these spaces over all exact
patterns gives the full Hilbert space \(\mathcal H_L\).

\paragraph{Counting the orthogonal sectors and their dimensions.}
Begin with a single active segment of length \(m\). In the basis
\eqref{eq:F-core-basis}, every allowed suffix \(t_2\cdots t_m\) fixes
one and only one first ket. Thus \(d_m=\dim\mathcal A_m\) is the number
of words of length \(m-1\) that avoid \(21\) and end in \(0\) or \(1\).
To extend a word by one letter, we only need to remember whether it
ends in \(2\), since precisely those words cannot be followed by \(1\).
This is the two-state transfer-matrix method of enumerative
combinatorics~\cite[Sec.~4.7]{StanleyEC}. Applying it to these suffixes,
as detailed in Appendix~\ref{app:F-count}, gives
\begin{equation}
\begin{aligned}
&d_m=F_{2m-1},\\
 &d_2=2,\quad d_3=5,\quad d_4=13,\quad d_5=34,
\end{aligned}
\label{eq:F-core-dim}
\end{equation}
where the \(F_k\) are the Fibonacci numbers, defined by \(F_0=0\),
\(F_1=1\), and \(F_{k+2}=F_{k+1}+F_k\) for \(k\ge0\).
The frozen ends in \eqref{eq:F-active} each contribute a single ket.
Likewise, a fixed \(21\) barrier contributes one ket. Hence a sector
\(\mathcal S\) of the full chain with active segments of lengths
\(m_1,\ldots,m_q\) has dimension
\begin{equation}
 \dim\mathcal S=\prod_{i=1}^{q}F_{2m_i-1}.
 \label{eq:F-sector-dim}
\end{equation}
The empty product is \(1\) for a fully frozen sector.

As a worked example of the full-chain count, take \(L=3\) and sort
the sectors by the position of the \(21\) barrier. If there is no barrier, \eqref{eq:F-frozen-ends} gives four frozen lines and
\eqref{eq:F-frozen} gives six. The three doublets are
\(\ket1\otimes\mathcal A_2\), \(\ket z\otimes\mathcal A_2\), and
\(\mathcal A_2\otimes\ket2\). The remaining sector is \(\mathcal A_3\)
itself. A barrier on bond \((1,2)\) leaves a free end spin, with three
frozen basis kets. The same holds for a barrier on \((2,3)\). The two
barrier positions are mutually exclusive. The sector count and the
dimension sum are listed below.
\begin{center}
\small\setlength{\tabcolsep}{2.7pt}
\begin{tabular}{@{}lrrrr@{}}
\hline
\shortstack[l]{Barrier\\pattern} & \shortstack{Frozen\\lines} & Doublets
 & \shortstack{Dim.-5\\sectors} & \shortstack{Dimension\\sum}\\
\hline
No barrier & 10 & 3 & 1 & 21\\
\(21\) on \((1,2)\) & 3 & 0 & 0 & 3\\
\(21\) on \((2,3)\) & 3 & 0 & 0 & 3\\
\hline
Total & 16 & 3 & 1 & 27\\
\hline
\end{tabular}
\end{center}
Thus \(N_3=16+3+1=20\) sectors account for \(16+3\cdot2+5=27\) states.
For general \(L\), assembling the intervals between barriers gives
(Appendix~\ref{app:F-count})
\begin{equation}
\begin{aligned}
\sum_{L\ge0}N_Lx^L&=\frac{1-x+x^2}{1-4x+5x^2-3x^3},\\
 \sum_{L\ge0}x^L\sum_{\mathcal S\ {\rm at}\ L}\dim\mathcal S&=\frac{1}{1-3x}.
\end{aligned}
\label{eq:F-count-results}
\end{equation}
The coefficient of \(x^L\) in the first series counts the \(N_L\) orthogonal
sectors of a length-\(L\) chain, each with weight one. In the second it
counts basis states, each sector contributing \(\dim\mathcal S\). Their
dimensions sum to \(3^L\), so by orthogonality the sectors span \(\mathcal H_L\).

For \(L\ge2\), the largest sector is \(\mathcal A_L\), whose active segment
occupies the entire chain. To see why splitting an active segment cannot
increase the dimension, take an allowed suffix
from each of \(\mathcal A_r\) and \(\mathcal A_s\). Joining them with a
single \(0\) gives an allowed suffix of length \(r+s-1\), and its two
original pieces can be recovered at their fixed lengths. Thus
\(d_rd_s\le d_{r+s}\). Iterating and using that \(d_m\) increases with
\(m\) bounds \eqref{eq:F-sector-dim} by \(d_L\). The sector \(\mathcal A_L\)
attains the bound. Its fraction of the Hilbert space is
\begin{equation}
\begin{aligned}
\frac{D_{\max}(L)}{3^L}&=\frac{F_{2L-1}}{3^L}
 \sim \frac{1}{\sqrt5\,\varphi}\left(\frac{\varphi^2}{3}\right)^L\longrightarrow0,\\
 \varphi&=\frac{1+\sqrt5}{2}.
\end{aligned}
\label{eq:F-strong}
\end{equation}
Thus \(H_{\mathrm{str}}\) exhibits strong fragmentation relative to the full
spin Hilbert space. Its largest irreducible sector occupies an
exponentially small fraction at large \(L\).

The conserved projectors of \(H_{\mathrm{str}}\) can now be easily written down.
Each sector \(\mathcal S\) in the decomposition above has a conserved
orthogonal projector \(Q_{\mathcal S}\). To construct it, write
\(P_w=\ket w\bra w\) for a normalized ket. A frozen interval in the
configuration \(w\) contributes \(P_w\), and a fixed barrier contributes
\(P_{21}\). For an interval of the form \eqref{eq:F-active}, the factor is
\begin{equation}
 P_1^{\otimes a}\otimes P_z^{\otimes b}\otimes P_{\mathcal A_m}
     \otimes P_2^{\otimes c},\qquad
 P_{\mathcal A_m}=\sum_t P_{\chi(t_2)}\otimes P_t ,
 \label{eq:F-interval-projector}
\end{equation}
where \(t=t_2\cdots t_m\) ranges over the suffixes in the orthonormal
basis \eqref{eq:F-core-basis}. For a two-site example of this projector,
the two allowed suffixes give \(P_{\mathcal A_2}=P_{v0}+P_{01}\). Taking the tensor product of the
interval and barrier projectors in their order along the chain gives
\(Q_{\mathcal S}\). Every \(h_j\) and \(h_j^\dagger\) preserves both
\(\mathcal S\) and its orthogonal complement, so
\([Q_{\mathcal S},h_j]=[Q_{\mathcal S},h_j^\dagger]=0\) for every bond.
Thus \(Q_{\mathcal S}\in\mathfrak C_{\mathrm{str},L}\) and
\([Q_{\mathcal S},H_{\mathrm{str}}]=0\). These projectors resolve the identity,
\begin{equation}
 Q_{\mathcal S}Q_{\mathcal S'}=\delta_{\mathcal S,\mathcal S'}Q_{\mathcal S},
 \qquad \sum_{\mathcal S}Q_{\mathcal S}=\id_{\mathcal H_L}.
 \label{eq:F-projector-resolution}
\end{equation}
Summing over just the sectors with a given barrier pattern \(S\) recovers
\(\Pi_S\) in \eqref{eq:F-pattern}. The individual \(Q_{\mathcal S}\) also
resolve the frozen configurations and active segments within that pattern.

\subsection{Fragmentation of \texorpdfstring{\(H_{\mathrm{wk}}\)}{H\textunderscore wk}}\label{sec:M-sectors}

Write \(h_j=h_{\mathrm{wk},j}\) in this subsection and keep the normalized
one-site kets \(\ket v,\ket z\) of \eqref{eq:F-vz}. The nonzero actions
on the computational basis are
\begin{equation}
\begin{aligned}
h_{\mathrm{wk}}\ket{00}&=\alpha\ket{01},\\
h_{\mathrm{wk}}\ket{02}&=\beta\ket{01},\\
h_{\mathrm{wk}}^\dagger\ket{01}&=\overline\alpha\ket{00}+\overline\beta\ket{02}.
\end{aligned}
\label{eq:M-moves}
\end{equation}
All three kets have a \(0\) on the left. To see what happens on the
right, put \(P_0=\ket0\bra0\), \(E=\ket1\bra v\), and
\(K=\operatorname{span}\{\ket1,\ket v\}\). Then
\begin{equation}
\begin{aligned}
h_{\mathrm{wk}}&=N P_0\otimes E,\qquad
 h_{\mathrm{wk}}^\dagger=N P_0\otimes E^\dagger,\\
 W&=K\oplus\mathbb C\ket z .
\end{aligned}
\label{eq:M-controlled}
\end{equation}
A left spin in \(\ket0\) enables the transition
\(\ket v\leftrightarrow\ket1\) on its right neighbor and stays in
\(\ket0\) throughout that move. A left spin in
\(\operatorname{span}\{\ket1,\ket2\}\) disables the bond in both
directions. On the right leg, \(\ket z\) is annihilated by both
\(E\) and \(E^\dagger\).

We now proceed to constructing the orthogonal sectors. First we show that a fixed \(21\) no longer cuts the chain.
The bond on the pair \(21\) still annihilates it in both directions.
To see how its left neighbor can change the \(2\), apply the local
rule \eqref{eq:M-controlled} to the three-site example \(\ket{021}\). This gives
\[
 h_1\ket{021}=\beta\ket{011}.
\]
The projector \(B_2\) of \eqref{eq:F-barrier} fixes the input and
annihilates the output. It therefore fails to commute with \(h_1\).
To find the orthogonal sectors we instead start at the two open ends,
where the left and right legs of \eqref{eq:M-controlled} play different
roles.

We will now find the first spin from the left that will begin an active segment.
At site 1 choose the computational basis. If this spin is \(1\) or
\(2\), both bond operators on \((1,2)\) vanish, and site 2 is now the
left end of the remaining chain. Repeating this step gives a frozen
prefix \(\ket w\), where every letter of \(w\) belongs to \(\{1,2\}\).
If the prefix fills the chain, \(\ket w\) spans a frozen line.
Otherwise call the first site in \(\ket0\) site \(i\), so that
\(w\) has length \(i-1\). Its left bond is absent or disabled by the
prefix. Its right bond acts on it through \(P_0\). Site \(i\) thus
remains in \(\ket0\). If \(i<L\), the bond on \((i,i+1)\) can act
freely between \(\ket v\) and \(\ket1\) at site \(i+1\).

Next we will remove the frozen suffix from the right end of the chain.
The last spin occurs only on the right leg of a bond, so its orthogonal
splitting \(W=K\oplus\mathbb C\ket z\) is preserved. On the
\(\ket z\) branch even this last bond vanishes. Remove that spin and
apply the same split at the newly exposed right end, stopping before
site \(i\). This gives a suffix of consecutive \(\ket z\)'s.
Its position at the edge matters. Inside the chain a \(\ket z\) on
the left leg of a bond has a nonzero \(\ket0\) component. For example,
\(h_{\mathrm{wk}}^\dagger\ket{z1}=\beta\ket{0v}\).
Thus an interior \(z\) can participate in a move.

\begin{figure*}[t]
\centering
\begin{tikzpicture}[
 x=1.46cm,y=1cm,
 title/.style={anchor=west,font=\small\sffamily\bfseries,inner sep=0pt},
 fixed/.style={draw=HFgray,fill=black!4,rounded corners=.6mm,
               minimum width=9mm,minimum height=6.5mm,inner sep=1pt},
 moving/.style={draw=HFact,fill=HFact!9,text=HFact,rounded corners=.9mm,
               minimum width=9mm,minimum height=6.5mm,inner sep=1pt},
 desc/.style={font=\footnotesize\sffamily,inner sep=1pt}]
 \node[title] at (0,0) {(a)\enspace An active interval on sites 3--7};
 \foreach \x/\s in {.5/1,1.5/2,2.5/0,7.5/z,8.5/z}
    \node[fixed] at (\x,-.54) {$\ket{\s}$};
 \foreach \x/\s in {3.5/W,4.5/W,5.5/W,6.5/K}
    \node[moving] at (\x,-.54) {$\s$};
 \node[desc] at (1,-1.08) {frozen prefix};
 \node[desc] at (2.5,-1.08) {first $0$};
 \node[desc] at (4.5,-1.08) {unrestricted interior};
 \node[desc] at (6.5,-1.08) {right end};
 \node[desc] at (8,-1.08) {frozen suffix};
 \node[title] at (0,-1.78) {(b)\enspace A frozen line with the same first $0$};
 \foreach \x/\s in {.5/1,1.5/2,2.5/0,3.5/z,4.5/z,5.5/z,6.5/z,7.5/z,8.5/z}
    \node[fixed] at (\x,-2.32) {$\ket{\s}$};
\end{tikzpicture}
\caption{An active sector and a frozen line of \(H_{\mathrm{wk}}\) on a length 9 open chain.}
\label{fig:M-sectors}
\end{figure*}
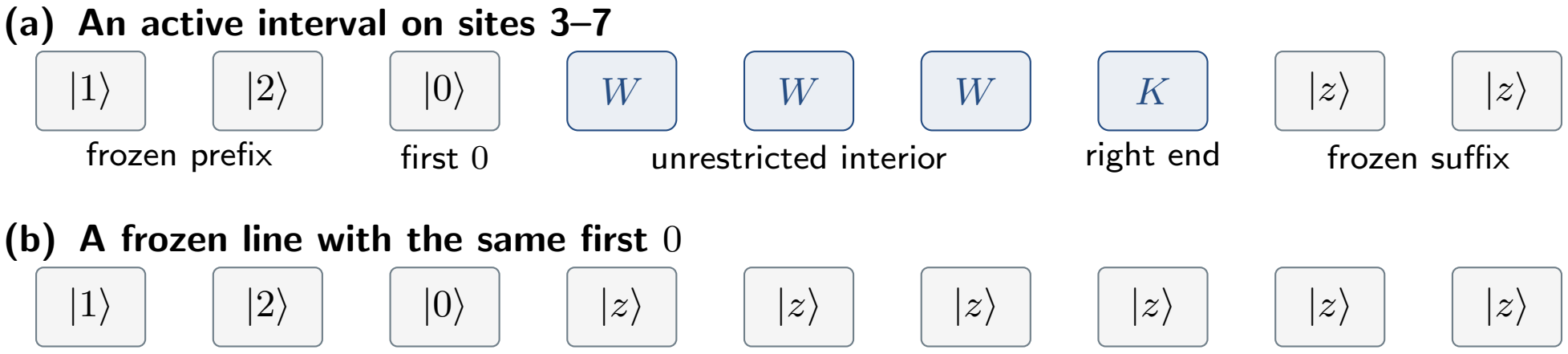

If every spin to the right of \(i\) has been removed, the result is
another frozen line. Otherwise let \(j>i\) be the first site from the
right selected in \(K\). The orthogonal sectors obtained by the two
end splittings have the following forms.
\begin{align}
 &\mathbb C\ket w,\qquad w\in\{1,2\}^L,
       \label{eq:M-frozen-prefix}\\
 &\mathbb C\bigl(\ket w\ket0\ket z^{\otimes(L-i)}\bigr),
       \label{eq:M-frozen}\\[-2pt]
 &\qquad w\in\{1,2\}^{i-1},\quad 1\le i\le L,\nonumber\\
 &\mathcal S_{w;i,j}=\mathbb C\ket w\otimes\mathcal M_{j-i+1}
                      \otimes(\mathbb C\ket z)^{\otimes(L-j)},
       \label{eq:M-sector}\\[-2pt]
 &\qquad w\in\{1,2\}^{i-1},\quad 1\le i<j\le L,\nonumber\\
 &\mathcal M_m=\mathbb C\ket0\otimes W^{\otimes(m-2)}\otimes K,\quad m\ge2.
       \label{eq:M-active}
\end{align}
Empty prefixes and suffixes are allowed. Equations
\eqref{eq:M-frozen-prefix} and \eqref{eq:M-frozen} give an orthonormal
product-state choice in the common kernel. Equation~\eqref{eq:M-sector}
describes a sector of the full chain. The frozen word \(w\) occupies
sites \(1,\ldots,i-1\), the active interval occupies sites \(i,\ldots,j\),
and every site after \(j\) is fixed in \(\ket z\).
Equation~\eqref{eq:M-active} gives the space on that interval. Its length
\(m=j-i+1\) includes both endpoints. Site \(i\) stays in \(\ket0\) and
enables the first move. Site \(j\) can move between \(\ket v\) and \(\ket1\)
in \(K\). The \(m-2=j-i-1\) sites strictly between them each have the full
spin-1 space \(W\). Figure~\ref{fig:M-sectors} illustrates the endpoint
construction on a nine-site chain. In panel (a), \(w=12\), \(i=3\) and \(j=7\). Sites 4--6 supply three factors of \(W\), so this sector has
dimension \(3^3\cdot2=54\). Panel (b) shows the frozen line obtained by
choosing \(\ket z\) on every site to the right of the same first \(0\).

We will now show that this active interval is a single sector.
Before giving the general proof, consider the shortest active interval
with an interior spin, \(m=3\). This example shows why the middle spin
in \eqref{eq:M-active} needs all of \(W\). With site 1 fixed in \(\ket0\), the first bond connects \(\ket1\)
and \(\ket v\) on site 2. Acting from the right now gives
\[
 h_2^\dagger\ket{0v1}=\overline\alpha\ket{00v},\qquad
 \ket0=\frac{\alpha}{N}\ket v+\frac{\overline\beta}{N}\ket z .
\]
The middle spin has acquired a \(z\) component. This explicitly shows
how a split that was conserved at the right edge ceases to be conserved
on an interior site. To rule out every possible further splitting, we
show by induction that every operator commuting with the restricted
bonds and adjoints on \(\mathcal M_m\) is a scalar. For the base case
of the induction take \(m=2\). The fixed left \(0\) leaves just \(K\) on the right. On \(K\),
\(E,E^\dagger,EE^\dagger,E^\dagger E\) are the four matrix units
in the basis \(\ket1,\ket v\), so their common commutant is scalar.

For \(m\ge3\), suppress the fixed first \(0\) and call the space to
the right of site 2 \(R=W^{\otimes(m-3)}\otimes K\). The first
bond acts on site 2 as \(E\), with its adjoint acting as
\(E^\dagger\). They generate all matrices on \(K\) and annihilate
\(\ket z\). Consequently an operator \(Q\) commuting with both has
the form
\begin{equation}
\begin{aligned}
Q&=P_K\otimes X+P_z\otimes Y,\\
 P_K&=\ket1\bra1+\ket v\bra v,\qquad P_z=\ket z\bra z,
\end{aligned}
\label{eq:M-commutant-step}
\end{equation}
where \(X,Y\) act on \(R\). In particular, \(Q\) cannot mix the
two summands because it commutes with
\(P_K=EE^\dagger+E^\dagger E\) on site 2.

The next bond couples those summands. After division by \(N\), it
acts as \(P_0\otimes E_3\), where \(E_3\) acts on the first site of
\(R\). Take the \(\ket v\) and \(\ket z\) diagonal matrix elements
on site 2 of \([Q,P_0\otimes E_3]=0\). Since
\(\langle v|P_0|v\rangle=|\alpha|^2/N^2\) and
\(\langle z|P_0|z\rangle=|\beta|^2/N^2\) are both nonzero, this gives
\([X,E_3]=[Y,E_3]=0\). The adjoint bond gives the same relations for
\(E_3^\dagger\). All bonds farther to the right act entirely on \(R\),
so \(X\) and \(Y\) commute with them and their adjoints as well.
These are exactly the generators on an active interval of length
\(m-1\) with its fixed initial \(0\) suppressed. By induction,
\(X=x\id_R\) and \(Y=y\id_R\).
Finally, the off-diagonal matrix element on site 2 reads
\[
 0=\langle v|[Q,P_0\otimes E_3]|z\rangle
   =(x-y)\langle v|P_0|z\rangle E_3 .
\]
The overlap is \(\alpha\beta/N^2\ne0\), so \(x=y\). This proves
that the restricted commutant is scalar for every \(m\). Its only
orthogonal projectors are zero and the identity. In particular the
\(K\) and \(z\) components of an interior spin belong to the same
irreducible sector, even though they label different choices at the
open right end.

\paragraph{Counting the sectors and their dimensions.}
There are \(2^L\) frozen words with no enabling spin. For a first
\(0\) at \(i\), each of the \(2^{i-1}\) prefixes gives one further
frozen line when the whole tail is \(z\). The number of frozen lines in
this decomposition is therefore
\begin{equation}
 f_L^{\mathrm{wk}}=2^L+\sum_{i=1}^L2^{i-1}=2^{L+1}-1 .
 \label{eq:M-frozen-count}
\end{equation}
An active interval of length \(m\) has one fixed site, \(m-2\) full
sites and a final doublet. Its left endpoint can be any
\(i=1,\ldots,L-m+1\). Once the prefix is chosen, both its right
endpoint \(j=i+m-1\) and the frozen suffix are fixed. Thus its
dimension \(d_m^{\mathrm{wk}}\) and multiplicity \(c_{L,m}^{\mathrm{wk}}\) are
\begin{equation}
\begin{aligned}
d_m^{\mathrm{wk}}&=2\cdot 3^{m-2},\qquad 2\le m\le L,\\
 c_{L,m}^{\mathrm{wk}}&=\sum_{i=1}^{L-m+1}2^{i-1}=2^{L-m+1}-1 .
\end{aligned}
\label{eq:M-multiplicity}
\end{equation}
Counting each sector once gives
\begin{equation}
 N_L^{\mathrm{wk}}=f_L^{\mathrm{wk}}+\sum_{m=2}^L c_{L,m}^{\mathrm{wk}}
            =3\cdot2^L-L-2 .
 \label{eq:M-count}
\end{equation}
For a direct comparison with the three-site count in
Section~\ref{sec:F-sectors}, apply the endpoint construction to
\(H_{\mathrm{wk}}\) at \(L=3\). The frozen lines number \(8+4+2+1=15\).
There are three doublets, \(\mathcal M_2\otimes\ket z\),
\(\ket1\otimes\mathcal M_2\), and \(\ket2\otimes\mathcal M_2\),
and one six-dimensional sector \(\mathcal M_3=\mathbb C\ket0\otimes W\otimes K\).
These \(19\) sectors account for \(15+3\cdot2+6=27\) states.

To verify completeness for every \(L\), hold a single prefix and its
first \(0\) at \(i\) fixed. There is one all-\(z\) tail. A tail with
right endpoint \(j\) contributes \(2\cdot 3^{j-i-1}\) states, so all
tail dimensions sum to
\[
 1+\sum_{j=i+1}^L 2\cdot 3^{j-i-1}=3^{L-i}.
\]
Restoring the prefix choices and the words without an enabling spin
gives the full dimension,
\begin{equation}
 \sum_{\mathcal S}\dim\mathcal S
    =2^L+\sum_{i=1}^L2^{i-1}\cdot 3^{L-i}
    =2^L+(3^L-2^L)=3^L .
 \label{eq:M-complete}
\end{equation}
The largest sector occurs at \(i=1,j=L\) and occupies a finite fraction
of the full Hilbert space at every length,
\begin{equation}
 D_{\max}^{\mathrm{wk}}(L)=2\cdot 3^{L-2},\qquad
 \frac{D_{\max}^{\mathrm{wk}}(L)}{3^L}=\frac29,\qquad L\ge2 .
 \label{eq:M-largest}
\end{equation}
Thus \(H_{\mathrm{wk}}\) exhibits weak fragmentation relative to the full
Hilbert space. Its number of sectors grows exponentially, yet a single
sector retains two ninths of all states. For \(H_{\mathrm{str}}\) the
corresponding fraction decays exponentially by \eqref{eq:F-strong}.

The conserved projectors of \(H_{\mathrm{wk}}\) can be easily written down as before.
Write \(P_w=\ket w\bra w\) on the prefix, with the empty prefix
contributing the scalar one. The orthogonal projector onto
\(\mathcal S_{w;i,j}\) is
\begin{equation}
 Q_{w;i,j}
 =P_w\otimes P_0\otimes\id_W^{\otimes(j-i-1)}
        \otimes P_K\otimes P_z^{\otimes(L-j)} .
 \label{eq:M-projectors}
\end{equation}
The two frozen families contribute respectively \(P_w\) on the whole
chain and \(P_w\otimes P_0\otimes P_z^{\otimes(L-i)}\). Their
ranges reduce every bond and its adjoint by the endpoint construction.
These projectors therefore belong to \(\mathfrak C_{\mathrm{wk},L}\).
Distinct choices are orthogonal at the first site where the end
splittings differ, and \eqref{eq:M-complete} shows that the projectors
sum to \(\id_{\mathcal H_L}\).

\subsection{Fragmentation on a periodic chain}\label{sec:ring}

What changes when the two ends of the chain are joined? The periodic
Hamiltonian in \eqref{eq:periodic-H} adds the oriented bond
\((h_\sigma)_{L,1}\). The same \(G^{\otimes L}\) conjugates this bond
as well, so the two periodic Hamiltonians remain locally similar.
We now include all \(L\) bonds and their adjoints in the bond
\(*\)-algebra. The sites are labeled, and indices are read modulo \(L\).
For \(L=2\) the sum includes both orientations of the pair.

\paragraph{The strongly fragmented chain.}
Every occurrence of \(21\) still cuts the chain, including one across
sites \(L,1\). Fixing a single barrier leaves an open interval of
\(L-2\) sites. Fixing the exact positions of \(k\geq1\) barriers leaves
\(k\) barrier-free intervals, each with the sector decomposition of
Section~\ref{sec:F-sectors}. Only the ring with no barrier anywhere
requires a new analysis. The number of labeled basis words that avoid
\(21\) cyclically is \(C_L=\varphi^{2L}+\varphi^{-2L}\), where
\(\varphi=(1+\sqrt5)/2\). Within this space there are three independent
frozen states, \(\ket1^{\otimes L}\), \(\ket2^{\otimes L}\), and
\(\ket z^{\otimes L}\). Their orthogonal complement within the
barrier-free space is a single irreducible sector of dimension
\(C_L-3\). It is the largest sector on the ring
(Appendix~\ref{app:ring}).

\paragraph{The weakly fragmented chain.}
On the open chain, a fixed first \(\ket0\) enables its right neighbor
and the final active spin belongs to \(K\). On the ring, the closing
bond can also act on these former endpoints. To see what can still
freeze, use the one-site basis \(\{\ket1,\ket2,\ket z\}\), with \(\ket z\) defined in
\eqref{eq:F-vz}. Its vectors are linearly independent for
\(\alpha\beta\ne0\). By \eqref{eq:M-controlled}, a left spin in
\(\operatorname{span}\{\ket1,\ket2\}\) disables a bond, and a right
spin in \(\ket z\) is killed by both moves. Thus, in a frozen basis
word, a \(z\) must be followed by another \(z\). Going around the ring
leaves either a word composed entirely of \(1,2\) or the all-\(z\)
word. These states span a frozen space of dimension \(2^L+1\). Its
orthogonal complement is one irreducible sector, as shown in
Appendix~\ref{app:ring}. The weak ring therefore has \(2^L+1\)
one-dimensional frozen sectors in any orthonormal resolution of this
space, and a single active sector of dimension \(3^L-2^L-1\).

\begin{table}[tb]
\caption{Largest irreducible orthogonal sector for open and periodic
boundaries, for \(L\geq2\).}\label{tab:ring}
\centering
\setlength{\tabcolsep}{3pt}
\begin{tabular}{@{}lccc@{}}
\hline
 & \shortstack{Open\\\(D_{\max}\)} & \shortstack{Periodic\\\(D_{\max}\)}
 & \shortstack{Periodic limit\\\(D_{\max}/3^L\)} \\
\hline
\(H_{\mathrm{str}}\) & \(F_{2L-1}\) & \(\varphi^{2L}+\varphi^{-2L}-3\) & \(0\) \\
\(H_{\mathrm{wk}}\)  & \(2\cdot3^{L-2}\) & \(3^L-2^L-1\) & \(1\) \\
\hline
\end{tabular}
\end{table}

Table~\ref{tab:ring} separates the effect of the open ends from that
of a bulk barrier. In the strong chain the largest-sector fraction
decays as \((\varphi^2/3)^L\) on the ring. In the weak chain it grows
from the open-chain value \(2/9\) to a periodic fraction
\(1-(2/3)^L-3^{-L}\) that tends to one, even though the number of
frozen sectors still grows exponentially.

\section{Discussion}\label{sec:discussion}

We have shown that two open chain spin Hamiltonians can be related by a
nonunitary on-site similarity transformation even when one exhibits strong
Hilbert space fragmentation and the other weak fragmentation. In our
spin-1 example, the same transformation relates the double-row transfer
matrices of \(H_{\mathrm{str}}\) and \(H_{\mathrm{wk}}\), and the two
Hamiltonians have the same Jordan form at every chain length.
To see how these statements are compatible, consider what the local
similarity \(S_L=G^{\otimes L}\) does to a sector projector \(Q\).
Writing \(\widetilde Q=S_LQS_L^{-1}\), we have
\[
 [\widetilde Q,h_{\mathrm{wk},j}]
   =S_L[Q,h_{\mathrm{str},j}]S_L^{-1}.
\]
Thus the commutant of the bond densities alone is carried exactly from
one chain to the other. An orthogonal sector, however, is defined using
the fixed spin inner product. Its projector must also commute with every
adjoint density, as in \eqref{eq:sector-algebras}.

Taking the adjoint of the transformed bond gives
\begin{equation}
 h_{\mathrm{wk},j}^{\dagger}
    =(S_L^{-1})^{\dagger}h_{\mathrm{str},j}^{\dagger}S_L^{\dagger}.
 \label{eq:discussion-adjoint}
\end{equation}
For a unitary \(S_L\) this would equal
\(S_L h_{\mathrm{str},j}^{\dagger}S_L^{-1}\), so the commutation
relations with the adjoints would carry over as well. For the nonunitary
\(G\) in \eqref{eq:G}, the distinction can already be seen on one bond.
Take the conserved barrier projector \(B_j\) of \eqref{eq:F-barrier}.
Its image acts on that bond as
\(\widetilde B=(\ket2-t\ket0)\bra2\otimes\ket1\bra1\).
It is an oblique projector that commutes with every transformed density.
The adjoint of the density on the barrier bond itself gives
\[
 h_{\mathrm{wk}}^{\dagger}\widetilde B\ket{21}=-tN\ket{0v},
 \qquad
 \widetilde B h_{\mathrm{wk}}^{\dagger}\ket{21}=0 .
\]
Since \(t=\beta/\alpha\ne0\), the commutator of
\(h_{\mathrm{wk}}^{\dagger}\) with \(\widetilde B\) is nonzero.
Extending this calculation by the identity on all other sites shows
that, for every \(L\ge2\),
\begin{equation}
 S_L\mathfrak C_{\mathrm{str},L}S_L^{-1}
    \ne \mathfrak C_{\mathrm{wk},L}.
 \label{eq:discussion-commutants}
\end{equation}
Thus \(S_L\) fails to carry the bond \(*\)-algebra of \(H_{\mathrm{str}}\)
onto that of \(H_{\mathrm{wk}}\), since conjugating an algebra also
conjugates its commutant.

The discrete symmetries of the Yang--Baxter equation also preserve
orthogonal fragmentation~\cite{GYBEsym}. To see this, let \(Q\) be an
orthogonal sector projector. Exchanging the two sites of each bond,
\(h\mapsto PhP\), is implemented on the open chain by the unitary
operator \(\mathcal R_L\) that reverses the order of the sites. The
corresponding projector is \(\mathcal R_L Q\mathcal R_L^\dagger\).
For example, a \(21\) barrier becomes a \(12\) barrier on the reflected
bond. Under full transposition, \(h\mapsto h^{\mathsf T}\), the
corresponding projector is \(\overline Q\), where the bar denotes
complex conjugation in the spin basis. Indeed, complex conjugation
takes the pair \(h_j,h_j^\dagger\) to
\(\overline{h_j},h_j^{\mathsf T}\), precisely the adjoint and the
bond density of the transposed model. Both operations therefore
preserve the dimensions and multiplicities of all irreducible
orthogonal sectors. A permutation of the local spin labels has the
same property, since it is an on-site unitary, while a nonzero
rescaling of \(h\) leaves the bond \(*\)-algebra unchanged.

For the two Hamiltonians studied here, the compatible twists of
Appendix~\ref{app:lift-orbits} also preserve every sector projector.
The reason is particularly simple for a one-site twist. Both
densities factor as \(h=A\otimes B\), as shown in
Eq.~\eqref{eq:factors}. The compatibility condition
\([h,V\otimes Z]=0\) therefore gives
\[
  (VAV^{-1})\otimes(ZBZ^{-1})=A\otimes B .
\]
Equality of these nonzero simple tensors implies
\(VAV^{-1}=\lambda A\) and \(ZBZ^{-1}=\lambda^{-1}B\) for some
\(\lambda\ne0\). Twisting either site thus only rescales \(h\).
The compatible two-site twists considered in
Appendix~\ref{app:lift-orbits} have the same effect. Each density
has a product output ket and a product input covector, and the twist
compatibility conditions preserve both lines. Thus, for this pair
of chains, it is the homogeneous non-unitary similarity \(G\) of
Section~\ref{sec:local-similarity} that changes strong fragmentation
into weak fragmentation.

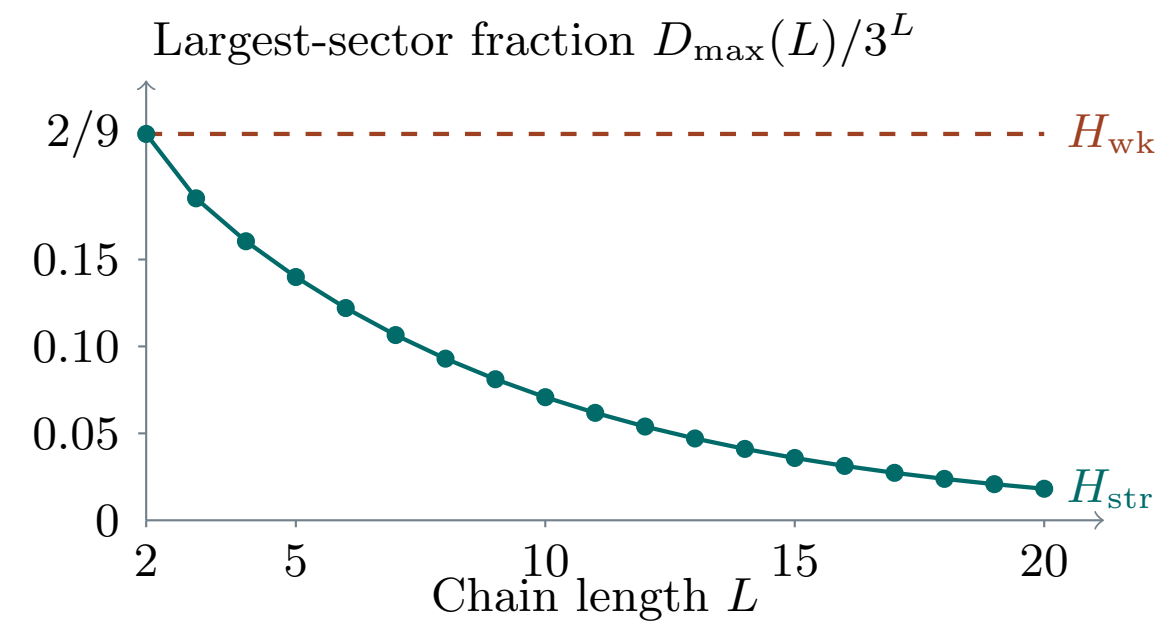
\begin{figure}[tb]
\centering
\begin{tikzpicture}[x=.33cm,y=11.5cm,font=\small]
  \foreach \y in {.05,.10,.15}{
    \draw[HFgray] (1.84,\y)--(2,\y);
  }
  \draw[HFgray,->] (2,0)--(21.2,0);
  \draw[HFgray,->] (2,0)--(2,.253);
  \node[anchor=south west,inner sep=1pt] at (2,.257)
      {Largest-sector fraction \(D_{\max}(L)/3^L\)};
  \foreach \l in {2,5,10,15,20}{
    \draw[HFgray] (\l,0)--(\l,-.004) node[below,text=black] {\(\l\)};
  }
  \node[anchor=north] at (11,-.023) {Chain length \(L\)};
  \foreach \y/\labeltext in {0/0,.05/0.05,.10/0.10,.15/0.15}{
    \node[anchor=east] at (1.82,\y) {\(\labeltext\)};
  }
  \node[anchor=east] at (1.82,{2/9}) {\(2/9\)};
  \draw[HFcut,thick,dashed] (2,{2/9})--(20,{2/9});
  \node[anchor=west,text=HFcut] at (20.1,{2/9}) {\(H_{\mathrm{wk}}\)};
  \draw[HFbar,thick] plot[mark=*,mark size=1.3pt] coordinates {
    (2,0.22222222) (3,0.18518519) (4,0.16049383) (5,0.13991770)
    (6,0.12208505) (7,0.10653864) (8,0.09297363) (9,0.08113601)
    (10,0.07080560) (11,0.06179049) (12,0.05392320) (13,0.04705759)
    (14,0.04106612) (15,0.03583750) (16,0.03127460) (17,0.02729266)
    (18,0.02381770) (19,0.02078518) (20,0.01813877)
  };
  \node[anchor=west,text=HFbar] at (20.1,.01813877) {\(H_{\mathrm{str}}\)};
\end{tikzpicture}
\caption{Fraction of the full Hilbert space occupied by the largest
orthogonal sector of each open chain for \(2\le L\le20\). The points
mark integer chain lengths for \(H_{\mathrm{str}}\). The dashed line
gives the constant fraction for \(H_{\mathrm{wk}}\).}
\label{fig:discussion-fractions}
\end{figure}

To quantify the difference in fragmentation, we compare the dimensions
of the largest orthogonal sectors of the two open chains at fixed \(L\).
In \(H_{\mathrm{str}}\), every occurrence of \(21\) cuts the chain
into independent intervals. In the largest sector the entire chain is active.
Its basis vectors are indexed by suffixes of length \((L-1)\) that avoid
\(21\) and end in \(0\) or \(1\). Counting them gives the Fibonacci
number \(F_{2L-1}\). In
\(H_{\mathrm{wk}}\), a bond immediately to the left can change the \(2\)
in \(21\). The largest sector fixes the first spin to \(\ket0\), allows
the last spin to range over the two-dimensional space \(K\), and leaves
all three states available at each of the \((L-2)\) interior sites.
Equations~\eqref{eq:F-strong} and~\eqref{eq:M-largest} therefore give
\begin{equation}
\begin{aligned}
\frac{D_{\max}^{\mathrm{str}}(L)}{3^L}&=\frac{F_{2L-1}}{3^L}\longrightarrow 0,\\
 \frac{D_{\max}^{\mathrm{wk}}(L)}{3^L}&=\frac{2\cdot 3^{L-2}}{3^L}=\frac29,\qquad L\ge2 .
\end{aligned}
\label{eq:discussion-fractions}
\end{equation}
Figure~\ref{fig:discussion-fractions} shows how the distinction emerges
with chain length. At \(L=2\) both largest sectors contain two of the
nine states. By \(L=20\), the largest sector of \(H_{\mathrm{str}}\)
occupies only \(1.81\%\) of the Hilbert space and its fraction continues
to fall exponentially. The largest sector of \(H_{\mathrm{wk}}\)
retains two ninths of all states at every length \(L\ge2\), even though
the number of sectors also grows exponentially in this chain. Closing
the chain removes both endpoint restrictions. Its active fraction then
tends to one, while the bulk barriers of \(H_{\mathrm{str}}\) continue
to enforce strong fragmentation (Section~\ref{sec:ring}).

Another natural question that arises is if there is a local way to physically distinguish the two Hamiltonians.
The fate of a \(21\) barrier gives such a local way to distinguish the two chains. Consider an open chain of three sites prepared in the state
\(\lvert021\rangle\). At time \(t\), we measure whether the last two
spins are still in the configuration \(21\), using the projector
\(P_{21}^{(23)}=\mathbf{1}\otimes\lvert21\rangle\langle21\rvert\).
In the strongly fragmented chain, \(H_{\mathrm{str}}\lvert021\rangle=0\),
so the initial state is stationary. In the weakly fragmented chain,
the first bond converts \(\lvert021\rangle\) into
\(\beta\lvert011\rangle\), and \(H_{\mathrm{wk}}\lvert011\rangle=0\).
The time evolution therefore terminates at first order in \(t\), giving
the exact unnormalized right state (with \(\hbar=1\))
\[
  e^{-itH_{\mathrm{wk}}}\lvert021\rangle
  = \lvert021\rangle-it\beta\lvert011\rangle .
\]
After normalizing the right state in the standard spin inner product,
the probability of observing \(21\) on the last two sites is
\[
  p_{\mathrm{str}}(t)=1,\qquad
  p_{\mathrm{wk}}(t)=\frac{1}{1+|\beta|^2t^2}.
\]
Thus the same preparation and the same local measurement distinguish
the two Hamiltonians for every \(t>0\).

Both these Hamiltonians are obtained from spectral \(R\)-matrices
constructed by Baxterizing two bond-wise lifts of the
constant, non-invertible \(4\times4\) solution \(R_{1,11}\)~\cite{Hietarinta,MPK}.
In the ordered qubit basis
\((\ket{00},\ket{01},\ket{10},\ket{11})\), the representative in
\eqref{eq:seed} reads
\[
 R_{1,11}=
 \begin{pmatrix}
 0&\alpha&\beta&0\\
 0&0&0&0\\
 0&0&0&0\\
 0&0&0&0
 \end{pmatrix}
 =\ket{00}\bigl(\alpha\bra{01}+\beta\bra{10}\bigr).
\]
The earlier construction Baxterized this solution and obtained
higher-dimensional spin chains by enlarging the representation of the
supersymmetry generators at each site~\cite[Secs.~4.8 and 5.1]{MPK}.
This is the \emph{site-wise} lifting procedure in the terminology of
Ref.~\cite{PK}. Here we combine the same Baxterization prescription
with the \emph{bond-wise} lifting method introduced there~\cite[Sec.~IV]{PK}.
The isometry \(J:(\mathbb C^2)^{\otimes2}\to W^{\otimes2}\) embeds the
entire qubit bond and gives the spin-1 density \(h_J=J R_{1,11}J^\dagger\).
Its square vanishes for every choice of \(J\), since \(J^\dagger J=\id_4\). To preserve the braid
relation, the images on overlapping bonds must also agree on the two
orders of three successive moves. For isometries that send basis kets
to distinct basis words, only three images enter the density,
\begin{equation}
\begin{aligned}
h_J&=\ket p\bigl(\alpha\bra{s_1}+\beta\bra{s_2}\bigr),\\
 (p,s_1,s_2)&=\bigl(J(00),J(01),J(10)\bigr),
\end{aligned}
\label{eq:discussion-word-lift}
\end{equation}
where \(J(ab)\) denotes the two-letter word labeling \(J\ket{ab}\).
The choices for \(p,s_1,s_2\) number \(9,8,7\), respectively. Of the
resulting \(504\) ordered triples, \(444\) preserve the braid relation.
For each of these, both three-move products vanish, and the linear
Baxterization gives \(\check R_J(u)=\id_9+u h_J\)~\cite{MPK}.
The derivative at zero recovers the lifted density. For
\((01,00,20)\) and \((01,00,02)\), the double-row construction
\eqref{eq:logder} then gives the open Hamiltonians \(H_{\mathrm{str}}\)
and \(H_{\mathrm{wk}}\), respectively. Appendix~\ref{app:lifts}
derives the count, including the \(60\) choices excluded by the
three-site test.

The distinction between fragmentation and the braid condition extends
to the whole family. Every lift has at least \(2^{L+1}-1\) independent
frozen states on the open chain, including lifts that fail the braid
test. The fraction occupied by the largest sector distinguishes the
braid-preserving lifts further. Among the \(444\) ordered choices it is
exponentially small in \(L\) for \(276\), of order \(L^{-1/2}\) for
\(96\), and bounded below by a positive constant for \(72\)
(Appendix~\ref{app:lift-frozen}). The pair studied here realizes the
first and third behaviors within a single Yang--Baxter equivalence class.

To place the class shared by \(H_{\mathrm{str}}\) and \(H_{\mathrm{wk}}\)
in the context of the full family, we classify the \(R\)-matrices of all
\(444\) braid-preserving lifts under homogeneous on-site similarity.
At fixed couplings and spectral normalization, this gives \(38\) classes
for generic \(\beta/\alpha\). If one also identifies
solutions under rescaling the spectral parameter, transposition and
exchange of the two sites, there are \(20\) generic classes.
Including compatible twists as in Ref.~\cite{GYBEsym}
further reduces it to just \(17\) classes. Appendix~\ref{app:lift-orbits} specifies the equivalences
and lists representatives. Our two matrices already belong to the same
class under homogeneous similarity alone, through the explicit \(G\)
in \eqref{eq:G}.

Within such a class, a local \emph{unitary} similarity does preserve the
orthogonal fragmentation. Suppose \(\check R'(u)=(U\otimes U)\check R(u)
(U\otimes U)^\dagger\) for a fixed one-site unitary \(U\). Differentiation
at zero conjugates every open-chain bond by \(U_L=U^{\otimes L}\).
Taking the adjoint conjugates each adjoint bond by the same \(U_L\).
The two bond \(*\)-algebras and their commutants therefore obey
\begin{equation}
 \mathfrak B'_L=U_L\mathfrak B_L U_L^\dagger,\qquad
 \mathfrak C'_L=U_L\mathfrak C_L U_L^\dagger ,
 \label{eq:discussion-unitary}
\end{equation}
where the primes label the second model. In particular,
\(Q\mapsto U_L Q U_L^\dagger\) bijects the minimal orthogonal sector
projectors and preserves their ranks. A product of on-site unitaries
also preserves the support of a local conserved projector, so a
barrier maps to a barrier on the same sites. All sector dimensions,
their multiplicities and the strong or weak fragmentation consequently
agree. The distinct largest-sector dimensions in \eqref{eq:discussion-fractions}
exclude any such local unitary between our two bond families.

\section{Outlook}\label{sec:outlook}

The on-site similarity relating \(H_{\mathrm{str}}\) to \(H_{\mathrm{wk}}\)
changes strong fragmentation into weak fragmentation while preserving the
spectrum at every chain length. How much freedom does a non-unitary
on-site similarity give us to tune the orthogonal sectors of a general
strongly fragmented Hamiltonian?
Two close precedents for this question are as follows. In a dipole-conserving spin chain,
adding four-site interactions can also change strong fragmentation into
weak fragmentation~\cite{SalaDipole}. In the strongly interacting
Hatano--Nelson chain, a similarity transformation within each fragment
explains the reality of the spectrum in the constrained
limit~\cite{GhoshFragmentation}. Can an on-site similarity remove fragmentation altogether,
even to the point that the transformed bond \(*\)-algebra acts
irreducibly on the full Hilbert space? The models studied in this work already give
an obstruction. On-site similarity preserves the rank and square-zero
property of each spin-1 bond density, so the argument of
Appendix~\ref{app:lift-frozen} still gives at least \(2^{L+1}-1\)
orthogonal frozen sectors on the open chain after any such transformation. It would be
worth determining which strongly fragmented Hamiltonians admit an
on-site similarity that removes orthogonal fragmentation completely.

The two open chains studied here exhibit both Hilbert space fragmentation
and Yang--Baxter integrability. Our earlier construction of integrable
non-Hermitian chains used site-wise lifts of the supersymmetry
generators~\cite[Sec.~5.1]{MPK}, whose local zero modes give a
site-wise mechanism for fragmentation~\cite{PK}. Bond-wise lifting
allows different mechanisms even for a single constant braid operator.
As Appendix~\ref{app:lifts} shows for \(R_{1,11}\), all \(504\) ordered
computational-basis lifts fragment, and the \(444\) that preserve the
braid relation already realize all three scalings of the largest-sector
fraction found in this family. Extending the construction to other
constant braid operators~\cite{Hietarinta} would let us ask which lifts
admit Baxterization and which fragmentation mechanisms their
Hamiltonians support.

The conserved orthogonal sector projectors commute with the double-row
transfer matrix at every spectral parameter, yet none of the nontrivial
projectors belongs to the algebra generated by its expansion coefficients.
Indeed, every nonzero bond action increases the number of spins in
\(\lvert1\rangle\) by one. In a basis ordered by this number,
Eq.~\eqref{eq:palindrome} therefore makes every element of the
transfer-matrix algebra triangular with a constant diagonal, so its only
idempotents are zero and the identity. These spin-1 chains thus show that
the commuting charges supplied by Yang--Baxter integrability need not
exhaust the symmetries responsible for Hilbert space fragmentation.
The scalene Yang--Baxter equation, a relation among three distinct
two-site operators, offers a possible route to these additional
symmetries~\cite{ScaleneHidden,ScaleneAffine,ScaleneAdditive}.
Can a scalene triple compatible with the open boundaries generate the
sector projectors themselves?

\begin{acknowledgments}
We used ChatGPT (OpenAI) for various analytical and computational checks, for literature search and for organizing and editing the manuscript. We are solely responsible for the results presented in this work.

\end{acknowledgments}
\appendix
\section{Counting the sectors of \texorpdfstring{\(H_{\mathrm{str}}\)}{H\textunderscore str}}
\label{app:F-count}

\paragraph{Counting states on one active segment.}
Let \(p_k\) count length-\(k\) words that avoid \(21\) and end in \(0\)
or \(1\), and \(q_k\) those that avoid \(21\) and end in \(2\).
Thus \(p_k+q_k\) counts all length-\(k\) words avoiding \(21\).
For an active segment on \(k+1\) sites, the suffix \(t_2\cdots t_{k+1}\)
has length \(k\) and must end in \(0\) or \(1\). Each such suffix fixes
the first ket uniquely by \eqref{eq:F-core-basis}, so \(d_{k+1}=p_k\)
for \(k\ge1\). To illustrate this count, take a three-site active
segment, whose suffix has length \(k=2\). The five suffixes
\(00,01,10,11,20\) give \(d_3=p_2=5\). The three words \(02,12,22\)
end in \(2\) and contribute to \(q_2\), giving \(p_2+q_2=8\).

The one-letter suffixes give the initial counts \(p_1=2\) and \(q_1=1\). Appending \(0\) is allowed
after all \(p_k+q_k\) words, while appending \(1\) is allowed only after the
\(p_k\) words. Appending \(2\) is always allowed. Hence
\begin{equation}
 p_{k+1}=2p_k+q_k,\qquad q_{k+1}=p_k+q_k .
 \label{eq:F-suffix-rec}
\end{equation}
Starting from \((2,1)=(F_3,F_2)\), the Fibonacci addition rule gives
\((p_k,q_k)=(F_{2k+1},F_{2k})\) by induction. Setting \(k=m-1\)
proves \eqref{eq:F-core-dim}.

To obtain the generating function for \(d_m\), first subtract the two
relations in \eqref{eq:F-suffix-rec}. Their difference gives
\(p_{k+1}-q_{k+1}=p_k\), so \(q_{k+1}=p_{k+1}-p_k\).
Substituting this into the first relation at the next step gives
\[
 p_{k+2}=2p_{k+1}+q_{k+1}
        =2p_{k+1}+(p_{k+1}-p_k)=3p_{k+1}-p_k .
\]
Since \(d_{k+1}=p_k\), setting \(m=k+3\) gives
\(d_m-3d_{m-1}+d_{m-2}=0\) for \(m\ge4\).

Now let \(G(x)=\sum_{m\ge2}d_m x^m\). Multiplying by \(x\) or \(x^2\)
shifts each term of \(G(x)\) up by one or two powers, respectively. Thus
\[
\begin{aligned}
 (1-3x+x^2)G(x)
 &=d_2x^2+(d_3-3d_2)x^3\\
 &\quad+\sum_{m\ge4}(d_m-3d_{m-1}+d_{m-2})x^m .
\end{aligned}
\]
Every term in the sum vanishes by the recurrence. The four-site
active segment illustrates the first cancellation. The coefficient of
\(x^4\) is \(13-3\cdot5+2=0\). The first two powers
survive because \(G(x)\) starts at \(x^2\). Inserting \(d_2=2\) and
\(d_3=5\) gives \(2x^2+(5-3\cdot2)x^3=2x^2-x^3\). Dividing by
\(1-3x+x^2\) now gives
\begin{equation}
 G(x)=\frac{2x^2-x^3}{1-3x+x^2}.
 \label{eq:F-core-series}
\end{equation}

\paragraph{Multiplicities in a barrier-free interval.}
Let the interval have \(n\) sites. For the frozen configurations
\eqref{eq:F-frozen-ends}, choosing \(a\in\{0,\ldots,n\}\) fixes
\(c=n-a\), giving \(n+1\) choices. For \eqref{eq:F-frozen}, one site
is already occupied by \(\ket0\), leaving \(a+b+c=n-1\). The nonnegative
solutions of \(a+b+c=k\) are the weak compositions of \(k\) into three
parts. The stars-and-bars argument~\cite[Sec.~1.2]{StanleyEC} represents
each solution by a row of \(k\) identical marks with two dividers.
The numbers of marks before, between
and after the dividers are \(a,b,c\). Choosing the two divider positions
out of \(k+2\) gives \(\binom{k+2}{2}\). Thus the number of frozen lines is
\[
 f_n=(n+1)+\binom{n+1}{2}=\binom{n+2}{2}.
\]
For an active segment of fixed length \(m\), \eqref{eq:F-active} leaves
\(a+b+c=n-m\). The sector dimension \(d_m\) therefore occurs with multiplicity
\begin{equation}
 c_{n,m}=\binom{n-m+2}{2},\qquad 2\le m\le n.
 \label{eq:F-multiplicity}
\end{equation}
For \(n=0\), \(f_0=1\) represents the empty interval and there is no
active segment. A four-site example is \(\ket{21}\otimes\ket{21}\),
where the two barriers leave an empty interval between them.

Let \(S(x)\) count the orthogonal sectors of one barrier-free interval.
Each sector on \(n\) sites contributes one term \(x^n\). Consider first
the frozen sectors. In the family \(\ket1^{\otimes a}\ket2^{\otimes c}\),
the two lengths \(a\) and \(c\) vary independently from zero upwards.
Each therefore contributes the geometric series \(1+x+x^2+\cdots\).
In the family \(\ket1^{\otimes a}\ket z^{\otimes b}\ket0\ket2^{\otimes c}\),
there are three independently variable lengths, and the fixed \(\ket0\)
occupies one additional site. Together the two families give
\[
 \begin{aligned}
 \sum_{n\ge0}f_nx^n
 &=(1+x+x^2+\cdots)^2+x(1+x+x^2+\cdots)^3\\
 &=\frac{1}{(1-x)^2}+\frac{x}{(1-x)^3}
 =\frac{1}{(1-x)^3}.
 \end{aligned}
\]

Now fix an active segment of length \(m\ge2\). In \eqref{eq:F-active}
the interval has length \(n=a+b+m+c\). Each choice of \(a,b,c\)
specifies one sector, regardless of how many basis states lie in
\(\mathcal A_m\). Thus the active segment contributes \(x^m\), and the
three frozen lengths again contribute a factor \((1-x)^{-3}\). This gives
\[
 \sum_{n\ge m}c_{n,m}x^n
 =\sum_{a,b,c\ge0}x^{a+b+m+c}
 =\frac{x^m}{(1-x)^3}.
\]
Adding the frozen contribution and summing over all active lengths
gives the interval series. Since \(\sum_{m\ge2}x^m=x^2/(1-x)\), we obtain
\begin{equation}
\begin{aligned}
S(x)&=\sum_{n\ge0}\left(f_n+\sum_{m=2}^n c_{n,m}\right)x^n\\
 &=\frac{1+\sum_{m\ge2}x^m}{(1-x)^3}
 =\frac{1-x+x^2}{(1-x)^4}.
\end{aligned}
\label{eq:F-interval-count}
\end{equation}
To see what \(S(x)\) counts, return to the three-site interval with
no barrier. Its sectors contribute to the coefficient of \(x^3\), which is
\(10+3+1=14\). There are ten frozen lines, three sectors with an active
segment of length two, and one with an active segment of length three.
To count states in those same sectors, a frozen line still has weight
\(1\), whereas a sector with an active segment of length \(m\) has weight
\(d_m\). This gives
\begin{equation}
\begin{aligned}
W(x)&=\sum_{n\ge0}\left(f_n+\sum_{m=2}^n c_{n,m}d_m\right)x^n\\
 &=\frac{1+G(x)}{(1-x)^3}=\frac{1}{1-3x+x^2},
\end{aligned}
\label{eq:F-interval-weight}
\end{equation}
where the last step uses \eqref{eq:F-core-series}. For the same
three-site example, \(W(x)\) weights each sector by its dimension. Its
coefficient of \(x^3\) is \(10+3\cdot2+1\cdot5=21\), as in
the first row of the three-site table.

\paragraph{Assembling the open chain.}
Exactly \(r\) fixed barriers occupy \(2r\) sites and leave \(r+1\)
ordered intervals of lengths \(n_0,\ldots,n_r\), with
\(n_0+\cdots+n_r=L-2r\). Empty intervals are allowed at either end
or between two barriers. The ordered lengths determine the barrier
positions uniquely. Conversely, every exact pattern
\eqref{eq:F-pattern} gives one such list, and each interval must itself
be free of \(21\). This is why concatenating intervals and fixed pairs
counts each sector once.

For a specified list of lengths, each interval \(i\) admits \(f_{n_i}\)
frozen lines and, for each \(2\le m_i\le n_i\), \(c_{n_i,m_i}\) sectors
with an active segment of length \(m_i\). Multiplying the appropriate multiplicities
counts the sectors with those selections. Multiplying the corresponding
dimensions gives \eqref{eq:F-sector-dim}. If different selections
produce the same dimension, their multiplicities add.

The open chain consists of one interval followed by an ordered sequence
of barrier--interval pairs. By the product and sequence rules for ordinary
generating functions~\cite[Theorem~I.1]{FlajoletSedgewick}, multiplication
sums over the interval lengths with the prescribed total. If \(I(x)\) is the
one-interval series, a chain with exactly \(r\) barriers contributes
\(x^{2r}I(x)^{r+1}\). With \(r=0\) this is \(I(x)\), and with \(r=1\)
it is \(I(x)\,x^2 I(x)\), one interval on either side of the fixed
pair. Summing the geometric series gives the full-chain series
\begin{equation}
 \sum_{r\ge0} x^{2r}I(x)^{r+1}=\frac{I(x)}{1-x^2 I(x)}.
 \label{eq:F-assembly}
\end{equation}
To count sectors, set \(I=S\). Substituting Eq.~\eqref{eq:F-interval-count}
into Eq.~\eqref{eq:F-assembly} gives the full-chain generating function
\[
\begin{aligned}
 N(x)&=\frac{S(x)}{1-x^2 S(x)}\\
 &=\frac{1-x+x^2}{(1-x)^4-x^2(1-x+x^2)} .
\end{aligned}
\]
Expanding the denominator gives \(1-4x+5x^2-3x^3\), since the two
\(x^4\) terms cancel. Clearing this denominator in the first generating
function of \eqref{eq:F-count-results} therefore gives
\[
 (1-4x+5x^2-3x^3)N(x)=1-x+x^2 .
\]
Matching powers of \(x\) first
gives \(N_0=1,\ N_1=3,\ N_2=8\), and then
\begin{equation}
 N_L=4N_{L-1}-5N_{L-2}+3N_{L-3}\qquad (L\ge3).
 \label{eq:F-number-recurrence}
\end{equation}
As examples of the recurrence, successive substitutions of
\(L=3,4,5\) give \(20,49,120\) sectors, respectively.
Finally, taking \(I=W\) weights every full-chain sector by the product
of its interval dimensions. Equations \eqref{eq:F-interval-weight}
and \eqref{eq:F-assembly} give
\begin{equation}
\begin{aligned}
\frac{W(x)}{1-x^2W(x)}&=\frac{1}{(1-3x+x^2)-x^2}\\
 &=\frac{1}{1-3x}=\sum_{L\ge0}3^Lx^L.
\end{aligned}
\label{eq:F-completeness}
\end{equation}
The dimension identity follows coefficient by coefficient, including
chains with any number of internal barriers.

\section{Sector dimensions and irreducibility on the periodic chain}
\label{app:ring}

\paragraph{Using the open-chain result at every cut.}
For a fixed model, omit the bond \((i-1,i)\) and let \(U_i\) be the
largest active sector of the resulting open chain, whose first site
is \(i\). Sections~\ref{sec:F-sectors} and~\ref{sec:M-sectors} give
\(U_i\) explicitly. In each open chain it is irreducible and occurs
with multiplicity one. Consequently an orthogonal projector \(Q\)
commuting with all the bonds of the ring and their adjoints preserves
\(U_i\) and restricts there to a scalar \(q_i\in\{0,1\}\). Indeed, its
component mapping \(U_i\) to any other open sector would intertwine
inequivalent irreducible representations, and hence vanishes.

The \(U_i\) for different cuts are nonorthogonal. In the strong chain,
\(U_i\) contains the product ket with \(v\) at site \(i\) and \(0\)
elsewhere, by \eqref{eq:F-core-basis}. In the weak chain it contains
the ket with \(v\) at site \(i-1\) and \(0\) elsewhere, by
\eqref{eq:M-active}. In either model the kets for distinct cuts have
nonzero overlap \(|\langle0|v\rangle|^2\). The eigenspaces of the
orthogonal projector \(Q\) for eigenvalues zero and one are orthogonal,
so all \(q_i\) coincide. Thus every periodic sector projector acts
as a single scalar on \(\sum_i U_i\). Once this sum is identified
as a reducing subspace, it follows that it is irreducible.

\paragraph{The barrier-free part of the strong ring.}
Let \(\mathcal H_0\) denote the span of the computational words having
no \(21\) on any cyclic bond. Since every barrier projector commutes
with the bonds and their adjoints, \(\mathcal H_0\) is reducing.
The local rules for a word in \(\mathcal H_0\) are encoded by
\begin{equation}
\begin{aligned}
M&=\begin{pmatrix}1&1&1\\1&1&1\\1&0&1\end{pmatrix},\\
 C_L&=\dim\mathcal H_0=\tr(M^L)=\varphi^{2L}+\varphi^{-2L}.
\end{aligned}
\label{eq:ring-cyclic-words}
\end{equation}
Here \(M_{ab}=1\) allows \(b\) to follow \(a\), with row and column
labels \(0,1,2\). The trace enforces the rule on the closing bond,
while site 1 remains distinguished. The eigenvalues of \(M\) are
\(0\) and \((3\pm\sqrt5)/2=\varphi^{\pm2}\).

For each cut, \(U_i\) lies in \(\mathcal H_0\) because its open-chain basis
already excludes \(21\), and its last spin, at \(i-1\), is \(0\) or
\(1\). We determine the orthogonal complement of \(\sum_i U_i\)
within \(\mathcal H_0\). Resolve each site into
\(\mathbb C\ket1\oplus V\), where \(V=\operatorname{span}\{\ket0,\ket2\}\).
The basis in \eqref{eq:F-core-basis} respects this resolution, so
orthogonality to every \(U_i\) can be tested separately for each
pattern of sites occupied by \(1\).

If all spins are \(1\), the sole ket \(\ket1^{\otimes L}\) is frozen.
If all spins lie in \(V\), the cut before \(i\) selects \(\ket0\)
at \(i-1\) and \(\ket v\) at \(i\), and leaves all other spins free
in \(V\). A vector orthogonal to every such space must be killed by
the contraction \(\bra0_{i-1}\bra v_i\) at every cut. Expand it in
the product basis built from \(\ket2,\ket z\) of \(V\). The only
ordered pair on which this contraction is nonzero is \(z2\), since
\(\langle0|2\rangle=\langle v|z\rangle=0\) and
\(\langle0|z\rangle\langle v|2\rangle\ne0\). Linear independence
on the remaining sites then rules out every word containing \(z2\).
On a ring the only remaining words are all \(2\) and all \(z\).
Both are frozen under every bond and its adjoint.

Finally, consider a mixed pattern. Choose a block of \(r\geq1\)
consecutive \(V\)-spins bordered on both sides by a \(1\), and number its sites
locally from 1 to \(r\). Its final spin must be \(0\), because \(21\)
is excluded. If \(r=1\), the cut before that spin includes the
entire pattern in its open active space. For \(r\geq2\), the
conditions imposed by the successive cuts within this block are
particularly simple. The cut before site 1 selects \(v\) on site 1.
A cut before an interior site \(k\), \(2\le k<r\), selects \(0\)
on \(k-1\) and \(v\) on \(k\). The cut before site \(r\) selects
\(0\) on \(r-1\), with the final spin already fixed to \(0\).
In each case the other sites range over the entire allowed pattern.
A vector orthogonal to the first cut must have \(z\) on site 1.
Since \(\langle0|z\rangle\ne0\), orthogonality at the next interior
cut forces \(z\) on site 2, and so on up to \(r-1\). Orthogonality
at the final cut now forces the vector to vanish. This excludes
every mixed pattern from the complement.

We have proved the orthogonal decomposition
\begin{equation}
 \mathcal H_0=\operatorname{span}\{\ket1^{\otimes L},
     \ket2^{\otimes L},\ket z^{\otimes L}\}
     \ \mathbin{\oplus}\ \sum_i U_i .
 \label{eq:ring-strong-active}
\end{equation}
The three displayed frozen kets are linearly independent. Since their
span is frozen and \(\mathcal H_0\) is reducing, the second summand
is reducing and hence irreducible by the cut argument. Its dimension
is \(C_L-3\). Fixing a barrier confines the remaining bonds to an open interval of
length \(L-2\). Its largest sector has dimension \(F_{2L-5}\) for
\(L\geq4\), and dimension one for an empty or single-site interval.
In either case the dimension is smaller than \(\dim U_i=F_{2L-1}\), which already
fits in the barrier-free active sector.

\paragraph{Counting all sectors in the strong ring.}
For an exact pattern with \(k\geq1\) barriers, choose one marked
barrier and read off the \(k\) successive gap lengths clockwise. They
sum to \(L-2k\), and a gap may be empty. The marked barrier starts at
one of \(L\) labeled sites, while each completed configuration has
\(k\) choices of mark. Each gap contributes the open barrier-free
interval series \(S(x)\) in \eqref{eq:F-interval-count} for counting
sectors, or \(W(x)\) in \eqref{eq:F-interval-weight} for counting
states. Writing \([x^n]\) for coefficient extraction gives
\begin{align}
 N_{L,\mathrm{per}}^{\mathrm{str}}
   &=4+\sum_{k=1}^{\lfloor L/2\rfloor}
       \frac{L}{k}[x^{L-2k}]S(x)^k,\nonumber\\
 3^L&=C_L+\sum_{k=1}^{\lfloor L/2\rfloor}
       \frac{L}{k}[x^{L-2k}]W(x)^k .
 \label{eq:ring-barrier-count}
\end{align}
The terms \(4\) and \(C_L\) count, respectively, the sectors and the
states in \eqref{eq:ring-strong-active}. For example, on a three-site
ring the barrier-free space consists of three frozen lines and one
15-dimensional sector. A barrier at any of the three bonds leaves
one free spin, giving three further frozen lines per bond. The
dimension sum is \(3+15+3\cdot3=27\).

\paragraph{The weak ring.}
Set \(D=\operatorname{span}\{\ket1,\ket2\}=\ker P_0\) and
\(Z=\mathbb C\ket z=K^\perp\). By \eqref{eq:M-controlled}, the
common kernel of a bond and its adjoint is
\((D\otimes W)+(W\otimes Z)\). Since \(W=D\oplus Z\) as a vector
space, this kernel is spanned by a subset of the product basis
\(\{\ket1,\ket2,\ket z\}^{\otimes2}\). We can impose all bond
constraints in that same basis even though \(D\) and \(Z\) are
not orthogonal. A \(z\) forces its right neighbor to be \(z\), so
the periodic frozen space is
\begin{equation}
 \mathcal F_{\mathrm{wk}}^{\rm per}=D^{\otimes L}+Z^{\otimes L},
 \qquad \dim\mathcal F_{\mathrm{wk}}^{\rm per}=2^L+1 .
 \label{eq:ring-weak-frozen}
\end{equation}

For the cut before \(i\), Eq.~\eqref{eq:M-active} fixes site \(i\)
to \(0\), puts site \(i-1\) in \(K\), and leaves the other sites
arbitrary. Hence \(U_i^\perp\) has the local condition
\((Z\otimes W)+(W\otimes D)\) on the ordered pair \((i-1,i)\).
This time a \(z\) forces its \emph{left} neighbor to be \(z\).
Following this rule around the ring gives the same space
\(D^{\otimes L}+Z^{\otimes L}\). Consequently
\((\sum_i U_i)^\perp=\mathcal F_{\mathrm{wk}}^{\rm per}\).
The complement of this frozen space is reducing, and the cut argument proves
that it is irreducible. Choosing an orthonormal basis in the frozen
space now gives \(2^L+2\) sectors in all, comprising \(2^L+1\) frozen lines
and the one active sector of dimension \(3^L-2^L-1\).

\section{Bond-wise lifts of the qubit braid operator}\label{app:lifts}

Every bond-wise isometric lift of \(R_{1,11}\) considered here fragments
the Hilbert space of the open spin-1 chain. This holds even when the
isometry maps basis states to superpositions, as the frozen-sector bound
in Appendix~\ref{app:lift-frozen} shows. Among the \(504\) ordered
computational-basis triples, fragmentation thus occurs both for the
\(444\) that preserve the braid relation and for the \(60\) that fail it.

\subsection{Counting the lifts that preserve the braid relation}
Fix \(\alpha\beta\ne0\). The finite enumeration uses computational-basis
images with no additional phases. To complete a triple in
\eqref{eq:discussion-word-lift} to an isometry one can send \(\ket{11}\)
to any of the six remaining basis kets. The seed and its adjoint both
annihilate \(\ket{11}\), so all six completions give the same density.
There are thus \(3024\) full ordered basis embeddings and \(504\) choices
of the three images that enter \(h_J\). At \(\alpha=\beta\) exchanging
the input words even gives the same matrix. Arbitrary bond isometries
can have superpositions as images and form a continuous family.

The same embedding can be written in the bond parasupersymmetry
construction of Ref.~\cite{PK}. In the ordered qubit bond basis
\(e_0=\ket{00},e_1=\ket{01},e_2=\ket{10},e_3=\ket{11}\), the
order-three parasupercharge is the shift
\(q=\sum_{a=0}^{2}e_a e_{a+1}^\dagger\).
Then \(q^4=0\), \(q^3(q^\dagger)^3=e_0 e_0^\dagger\), and
\[
 h_{1/2}=q^3(q^\dagger)^3(\alpha q+\beta q^2).
\]
Replacing \(q\) by \(JqJ^\dagger\) in this expression gives \(h_J\),
because \(J^\dagger J=\id_4\) in every product on a single bond.
This explains why the one-bond construction works for every isometry.
Preservation of the braid relation requires the following check on the
shared spin.

Write \(p=ij\), let \(S=\{s_1,s_2\}\), and set
\(A=h_J\otimes\id_3\), \(B=\id_3\otimes h_J\). Each nonzero move
replaces a member of \(S\) by \(ij\). Reading operator products from
right to left, the only possible nonzero paths are
\begin{equation}
\begin{aligned}
ABA:\quad s_a k&\longmapsto ijk\longmapsto iij\longmapsto ijj\\
 &\hspace{1em}\text{if }jk\in S\text{ and }ii\in S,\\
 BAB:\quad k s_a&\longmapsto kij\longmapsto ijj\longmapsto iij\\
 &\hspace{1em}\text{if }ki\in S\text{ and }jj\in S .
\end{aligned}
\label{eq:liftpaths}
\end{equation}
Here \(k\in\{0,1,2\}\) and \(s_a\) denotes either input. If \(i=j\),
both paths require the output itself to be an input. The three diagonal
outputs therefore admit all \(3\cdot8\cdot7=168\) ordered choices.
If \(i\ne j\), the two paths end in distinct words. Whenever either
path exists its amplitude is nonzero, so \(ABA=BAB\) requires both paths
to be absent. For each fixed output \(ij\), the first path excludes the
three unordered input pairs \(\{ii,jk\}\), and the second excludes the
three pairs \(\{ki,jj\}\). Their only common pair is \(\{ii,jj\}\).
Restoring the order of the two inputs gives \(2(3+3-1)=10\) failures
per off-diagonal output. The six such outputs account for all \(60\)
failures, leaving
\[
 3\cdot56+6(56-10)=168+276=444
\]
braid-preserving lifts. As a check, our two triples have \(p=01\) and
\(00\in S\), but neither has an input starting with \(1\) or the input
\(11\). Both paths are blocked. Changing the second input to \(10\)
allows \(ABA\ket{000}=\alpha^2\beta\ket{011}\), while
\(BAB\ket{000}=0\).

For every braid-preserving lift, put \(X_j(u)=\id+u h_{J,j}\). The difference
between the two sides of \eqref{eq:braid} is
\(uv(A^2-B^2)+uv(u+v)(ABA-BAB)\), which vanishes by the two tests above.
Thus \(\check R_J(0)=\id_9\), \(\check R_J(u)^{-1}=\id_9-u h_J\),
and the ordinary \(R_J(u)=P\check R_J(u)\) is regular.

For each lift \(J\) that preserves the braid relation, the open-chain
Hamiltonian \(H_J=\sum_{j=1}^{L-1}h_{J,j}\) is the coefficient of \(u\)
in a family \(T_J(u)\) of mutually commuting operators.
The additive braid relation, \(X_j(u)X_j(v)=X_j(u+v)\), and commutation
at distance give \([T_J(u),T_J(v)]=0\) for
\(T_J(u)=X_{L-1}(u)\cdots X_1(u)\)~\cite[Cor.~5.4]{FK}.
Since \(T_J(0)=\id\), its logarithmic derivative gives
\(\sum_{j=1}^{L-1}h_{J,j}\). For the two chains of Section~\ref{sec:two},
\eqref{eq:logder} also gives the corresponding double-row transfer matrices.

\subsection{Frozen states and the sizes of the largest sectors}\label{app:lift-frozen}
For any isometry \(J\), the lifted rank-one density can be written
\(h_J=N\ket{a}\bra{b}\) with \(\ket a,\ket b\) orthonormal. Its adjoint is $h_J^\dagger=N\ket{b}\bra{a}$. Thus, a frozen state is killed by both $h_J$ and $h_J^\dagger$ precisely when its  contractions with both \(\bra b\) and
\(\bra a\) vanish, respectively. We impose both conditions since the density need not be Hermitian. Denote the common kernel on \(L\) sites by \(\mathcal{F}_L\)
and its dimension by \(f_L\). Extend a frozen prefix of $L-1$ sites by one site. The two contractions with $\bra a$ and $\bra b$ on the last bond, respectively, define a combined linear map $$ \mathcal{C}_L: \mathcal{F}_{L-1}\otimes W \longrightarrow \mathcal{F}_{L-2}\oplus\mathcal{F}_{L-2}.$$
Both contractions take values in $\mathcal{F}_{L-2}$, because they act only on the last two sites, leaving the constraints on the first $(L-2)$ sites unchanged. The kernel of this combined map is precisely $\mathcal{F}_{L}$, since the constraints on the earlier bonds are already satisfied and the two contractions impose the remaining conditions on the last bond. Since the codomain has dimension $2 f_{L-2}$, rank-nullity gives
\[
 f_L\ge3f_{L-1}-2f_{L-2},\qquad f_0=1,\quad f_1=3 .
\]
It follows that \(f_L-f_{L-1}\ge2(f_{L-1}-f_{L-2})\ge2^L\).
Summing yields \(f_L\ge2^{L+1}-1\). Choose an orthonormal basis of
\(\mathcal F_L\). Each basis vector \(\ket{\psi}\) is annihilated by
every bond density and its adjoint, so its orthogonal projector
\(\ket{\psi}\bra{\psi}\) lies in the common bond commutant.
Every lift therefore has at least \(2^{L+1}-1\) mutually orthogonal
one-dimensional frozen sectors. This exponential lower bound establishes
Hilbert space fragmentation independently of the braid relation.

For comparison, the open-chain classification by largest-sector size
is given below. Entries count ordered triples, and \(r_L=D_{\max}(L)/3^L\).
The mechanism-by-mechanism derivation will be shown in ref.~\cite{QutritStudy}.
\begin{center}
\small\setlength{\tabcolsep}{3pt}
\begin{tabular}{@{}lrr@{}}
\hline
Behavior of \(r_L\) & Braid holds & Braid fails\\ \hline
Exponentially small & 276 & 36\\
Of order \(L^{-1/2}\) & 96 & 0\\
Positive lower bound & 72 & 24\\ \hline
Total & 444 & 60\\ \hline
\end{tabular}
\end{center}
Table~\ref{tab:mechanisms} gives one braid-preserving density for each
mechanism. Write \(n_j^{(a)}=(\ket a\bra a)_j\) for the projector onto
state \(a\) at site \(j\), and
\(\ket{\chi}=(\beta\ket0-\alpha\ket1)/\sqrt{|\alpha|^2+|\beta|^2}\).
\begin{table*}[t]
\caption{Representative braid-preserving densities, fragmentation mechanisms,
and largest-sector fractions on the open chain.}\label{tab:mechanisms}
\centering
\renewcommand{\arraystretch}{1.2}
\begin{tabular}{@{}lll@{}}
\hline
Bond density \(h_J\) & Fragmentation mechanism & \(r_L=D_{\max}(L)/3^L\)\\ \hline
\(\ket{00}(\alpha\bra{01}+\beta\bra{10})\)
 & Immobile one-site hole \(\ket2\) & Exponentially small\\
\(\ket{00}(\alpha\bra{11}+\beta\bra{22})\)
 & Conserved reduced string & Exponentially small\\
\(\ket{01}(\alpha\bra{00}+\beta\bra{20})\)
 & Immobile bond \(\ket{21}\) (\(H_{\mathrm{str}}\)) & Exponentially small\\
\(\ket{22}(\alpha\bra{00}+\beta\bra{01})\)
 & Frozen interval \(\ket{\chi}\otimes\ket{21}\) & Exponentially small\\ \hline
\(\ket{01}(\alpha\bra{02}+\beta\bra{10})\)
 & Uniform charge \(\sum_j n_j^{(0)}\) & \(\Theta(L^{-1/2})\)\\
\(\ket{00}(\alpha\bra{01}+\beta\bra{22})\)
 & Staggered charge \(\sum_j(-1)^j n_j^{(2)}\) & \(\Theta(L^{-1/2})\)\\ \hline
\(\ket{00}(\alpha\bra{01}+\beta\bra{12})\)
 & Macroscopic sector & \(7/27\) for \(L\ge3\)\\ \hline
\end{tabular}
\end{table*}
In the second example, repeatedly deleting adjacent equal letters gives
the same reduced string before and after any pair flip~\cite{PK,MM}.
In the fourth, destructive interference in \(\ket{\chi}\) freezes the
three-site interval for arbitrary states on the neighboring sites.
The four mechanisms with exponentially small \(r_L\) account for
\(48,36,144,48\) braid-preserving triples, respectively. For the two
charge examples the largest irreducible sector attains the order
\(3^L/\sqrt L\) of the largest charge eigenspace.

For the last example, let
\(K=\operatorname{span}\{\ket{00},\bar\alpha\ket{01}+\bar\beta\ket{12}\}\).
The sector generated by the states beginning with \(00\) is
\[
 \bigl(K\otimes W+\mathbb C\ket{012}\bigr)\otimes W^{\otimes(L-3)}.
\]
Acting on the first bond gives \(K\otimes W\) on three sites, and
\(h_{23}^\dagger\ket{000}=\bar\alpha\ket{001}+\bar\beta\ket{012}\)
adds a seventh direction. These seven directions span a reducing
subspace on three sites. The next bond preserves the occupation of
\(\ket2\) on site~3, so adjoining an arbitrary suffix preserves this
subspace. To check irreducibility, compress \(h_{23}\) and
\(h_{23}h_{12}^\dagger\) to the prefix \(00\). On site~3 they give
nonzero multiples of \(\ket0\bra1\) and \(\ket0\bra2\), which together
with their adjoints generate all matrix units. Sandwiching each successive
bond between matrix units on its left site gives the same two operators
on its right site. Thus the algebra contains all matrix units between
words beginning with \(00\), and their cyclic span is irreducible.
On its orthogonal complement the first bond and its adjoint vanish,
so every remaining sector has dimension at most \(D_{\max}(L-1)\).
Starting with \(D_{\max}(2)=2\), this gives \(D_{\max}(L)=7\cdot3^{L-3}\)
for \(L\ge3\).

\subsection{Equivalences of the lifted \texorpdfstring{\(R\)}{R}-matrices}\label{app:lift-orbits}
Write \(t=\beta/\alpha\) and fix the couplings in both matrices being
compared. The labels \(0,1,2\) denote the three basis states of a
single spin-1 site. A two-letter word \(ij\) labels the bond state
\(\ket{ij}=\ket i\otimes\ket j\). We write \(p/s_1/s_2\) for an
ordered triple of distinct two-site words. In \eqref{eq:discussion-word-lift},
\(p\) is the output, and \(s_1,s_2\) are the inputs with amplitudes
\(\alpha,\beta\), respectively. We call the triple \emph{braid-preserving}
if its density satisfies \(h_{12}h_{23}h_{12}=h_{23}h_{12}h_{23}\) on
three sites. Appendix~\ref{app:lifts} gives \(444\) such triples among
the \(9\cdot8\cdot7=504\) possible choices.

Since \(\check R(u)=\id_9+u h\), we can describe each symmetry
by its action on \(h\) in the braided convention. The two discrete
symmetries of the YBE used here are site exchange \(h\mapsto PhP\)
and full transposition \(h\mapsto h^{\mathsf T}\) in the computational
basis. Applying either operation twice returns \(h\)~\cite{Hietarinta,GYBEsym}.
An early study of discrete YBE symmetries in vertex models is
Ref.~\cite{BellonMaillardViallet}.

Homogeneous gauge transformations and scaling are continuous YBE
symmetries~\cite{Hietarinta,GYBEsym}. The former conjugate \(h\) by
\(S\otimes S\), with the same \(S\in\mathrm{GL}(3,\mathbb C)\) on
both sites. Scaling the constant braid solution \(h\mapsto c h\),
with \(c\ne0\), sends \(\check R(u)\) to \(\check R(cu)\).
Compatible twists form a further, constrained family of continuous
YBE symmetries~\cite{GYBEsym}. Each row of the following table adds
the indicated operation to those already allowed.
\begin{center}
\small\setlength{\tabcolsep}{2.5pt}
\begin{tabular}{@{}p{.52\columnwidth}rrr@{}}
\hline
Allowed operations & Generic \(t\) & \(t=1\) & \(t=-1\)\\ \hline
Homogeneous gauge at fixed \(u\) & 38 & 30 & 35\\
Also \(u\mapsto c u\) & 34 & 30 & 31\\
Also site exchange and full transposition & 20 & 18 & 19\\
Also compatible twists & 17 & 17 & 17\\ \hline
\end{tabular}
\end{center}
For the one-site twists in the last row, the compatibility condition is
\([h,V\otimes Z]=0\) for invertible \(V,Z\). One may then conjugate \(h\) by
\(V\otimes\id\) or by \(\id\otimes Z\)~\cite[Lemma~2.2]{GYBEsym}.
These one-site twists already produce the final 17 classes. For an
early construction of twists of quasitriangular Hopf algebras, see
Ref.~\cite{ReshetikhinTwist}.

To count the homogeneous gauge classes, first consider the basis
permutations. A permutation \(\pi\) of the labels \(0,1,2\) acts by
\(S_\pi\ket i=\ket{\pi(i)}\) on every site. Conjugation by
\(S_\pi\otimes S_\pi\) therefore replaces each label in all three
words by its image under the same permutation and preserves the braid
relation. For example, \(01/00/20\) for \(H_{\mathrm{str}}\) encodes
the moves \(\ket{00}\mapsto\alpha\ket{01}\) and
\(\ket{20}\mapsto\beta\ket{01}\). Exchanging the basis labels \(1\)
and \(2\) sends this triple to \(02/00/10\).

The three words in a triple must use at least two different labels,
since a single label \(i\) gives only the word \(ii\). A permutation
that leaves the ordered triple unchanged must therefore fix at least
two of the three labels and hence also the third. Only the identity
fixes the triple, so the \(3!=6\) basis permutations give six distinct
ordered triples in each orbit. This leaves \(444/6=74\) representatives
to compare under general homogeneous gauge transformations.

For a source \(ij/s_1/s_2\), encode the input covector by the
\(3\times3\) matrix \(C=E_{s_1}+t E_{s_2}\), where \(E_{ab}\) is a
matrix unit for the word \(ab\). Let \(pq\) and \(D\) denote the output
and input matrix of a candidate target. Equality of the rank-one
densities after conjugation requires, and is implied by,
\begin{equation}
\begin{aligned}
S e_i&=a e_p,\qquad S e_j=b e_q,\\
 S^{\mathsf T}D S&=ab\,C,\qquad \det S=1 .
\end{aligned}
\label{eq:liftgauge}
\end{equation}
Here \(e_i=\ket i\), and an overall rescaling of \(S\) allows the
determinant normalization. For example, the first two equations match
the ket factors, and the third matches the input bras with the same
overall factor. With free spectral rescaling, only proportionality of
\(S^{\mathsf T}DS\) and \(C\) is required.
Comparing each candidate with the existing representatives by this
polynomial test gives Table~\ref{tab:liftreps}. One can eliminate the
linear constraints on \(S\) first and test whether the remaining
polynomials generate the unit ideal. Over \(\mathbb C\) this happens
precisely when there is no invertible gauge satisfying the equations.
The generic homogeneous-gauge counts in the first two rows apply to
every \(t\ne0,\pm1\). At \(t=1\), even exchanging the two inputs gives
the identical matrix.

\begin{table*}[t]\centering
\caption{The 38 homogeneous gauge classes at fixed spectral coordinate
and generic \(t=\beta/\alpha\). For each class, the \emph{Triple} column gives
one representative \(p/s_1/s_2\), where \(p\) is the output word and
\(s_1,s_2\) are the input words with amplitudes \(\alpha,\beta\),
respectively. The \emph{Lifts} column counts all braid-preserving ordered
triples in that class, including the representative. Each triple is counted
once regardless of the choice of \(J\ket{11}\). For example, \(01/00/02\)
for \(H_{\mathrm{wk}}\) represents class~13. Its 24 triples include
\(01/00/20\) for \(H_{\mathrm{str}}\), whose density is related to the
representative by the on-site similarity in \eqref{eq:gaugeR}.
Classes with the same orbit letter become equivalent under the full set
of operations in Appendix~\ref{app:lift-orbits}.}\label{tab:liftreps}
\begin{tabular}{rlrc@{\qquad}rlrc}
\hline
Class & Triple & Lifts & Orbit & Class & Triple & Lifts & Orbit\\ \hline
1 & \(00/01/02\) & 30 & A & 20 & \(01/02/21\) & 12 & M \\
2 & \(00/01/10\) & 6 & B & 21 & \(01/02/22\) & 12 & N \\
3 & \(00/01/12\) & 12 & C & 22 & \(01/10/02\) & 6 & K \\
4 & \(00/01/20\) & 12 & D & 23 & \(01/10/12\) & 12 & O \\
5 & \(00/01/22\) & 12 & E & 24 & \(01/10/21\) & 6 & K \\
6 & \(00/10/01\) & 6 & B & 25 & \(01/10/22\) & 6 & P \\
7 & \(00/10/11\) & 30 & A & 26 & \(01/11/12\) & 24 & H \\
8 & \(00/10/21\) & 12 & C & 27 & \(01/11/22\) & 12 & J \\
9 & \(00/10/22\) & 12 & E & 28 & \(01/12/10\) & 12 & O \\
10 & \(00/11/12\) & 24 & F & 29 & \(01/12/20\) & 12 & P \\
11 & \(00/11/22\) & 6 & G & 30 & \(01/12/21\) & 6 & L \\
12 & \(00/12/21\) & 6 & G & 31 & \(01/12/22\) & 12 & Q \\
13 & \(01/00/02\) & 24 & H & 32 & \(01/20/02\) & 6 & L \\
14 & \(01/00/21\) & 12 & I & 33 & \(01/20/21\) & 12 & H \\
15 & \(01/00/22\) & 12 & J & 34 & \(01/20/22\) & 12 & Q \\
16 & \(01/02/10\) & 6 & K & 35 & \(01/21/10\) & 6 & K \\
17 & \(01/02/11\) & 12 & I & 36 & \(01/21/12\) & 6 & L \\
18 & \(01/02/12\) & 12 & H & 37 & \(01/21/22\) & 12 & N \\
19 & \(01/02/20\) & 6 & L & 38 & \(01/22/10\) & 6 & P \\
\hline
\end{tabular}
\end{table*}

The numbered classes also specify the exceptional coupling ratios.
At \(t=-1\), classes \((2,6),(19,32),(30,36)\) merge pairwise.
At \(t=1\) these mergers hold as well, together with
\((11,12),(16,22),(23,28),(24,35),(25,38)\). If rescaling \(u\) is
allowed, the four pairs \((16,22),(23,28),(24,35),(25,38)\) already
merge for generic \(t\). For instance, \(\operatorname{tr}(Ph)\) is
unchanged by homogeneous conjugation, since \(P\) commutes with
\(S\otimes S\). It equals \(\beta\) for class 16 and \(\alpha\) for
class 22, so the distinction is detectable at fixed spectral normalization.

For the third row of the count, compare also with the reflected and
transposed representatives. At generic \(t\), adding compatible twists
then identifies three further pairs of classes, represented by
\[
\begin{gathered}
 00/11/22\ \sim\ 00/12/21,\qquad
 01/10/22\ \sim\ 01/12/20,\\
 01/02/20\ \sim\ 01/20/02 .
\end{gathered}
\]
For the first two pairs one may swap the right-site letters \(1,2\)
or \(0,2\), respectively. In each case the compensating left-site
matrix fixes \(e_0\) and sends \(e_1\mapsto t^{-1}e_2\),
\(e_2\mapsto t e_1\), so that the product commutes with \(h\).
For the third pair take \(V=\mathrm{diag}(1,1,t^{-2})\) and
\(Z=\mathrm{diag}(t^2,1,1)\), followed by an overall rescaling.

One can check that further one-site twists leave the 17 lettered
classes distinct using only one-site lines and their pairings.
For a product input covector, record whether the two output lines
coincide, whether the two input lines coincide, and which of the four
input--output pairings vanish. A product stabilizer fixes each factor
line in its own slot, so either one-site twist only rescales the bond
in this case. For a rank-two input matrix \(C\), use the four lines
\(x=\mathbb C e_i\), \(y=\mathbb C e_j\),
\(\ell=\ker C^{\mathsf T}\), and \(r=\ker C\). Record their coincidences
and the vanishing of \(C(x,m)\) and \(C(m,y)\) on each marked line \(m\),
where \(C(v,w)=v^{\mathsf T}Cw\).
A right twist can equally be written as the inverse left twist since
\(V\otimes Z\) stabilizes \(h\). The right description fixes the row
radical and the column evaluated on \(y\) up to scale, and the left
description fixes the column radical and the row evaluated on \(x\) up
to scale. Both output lines also stay fixed. Thus these data are
invariant, up to the relabeling induced by reflection and transposition
(for rank two, place the product tensor on the output side).
Evaluating them on Table~\ref{tab:liftreps} gives a different signature
for each letter. For example, classes 11 and 13 have input matrices of different
ranks, and remain distinct even after the twists. Both chains in the
main text lie in class 13, as witnessed by \eqref{eq:gaugeR}.

The more general compatible two-site twists of
Ref.~\cite[Lemma~2.10]{GYBEsym} give the same 17 classes here.
To see this, write the twist as \(F\) and the new density as
\(FhF^{-1}\). The mixed commutator \([h_{12},F_{13}F_{23}]=0\)
implies that \(F_{13}F_{23}\) preserves \(\mathrm{im}(h_{12})\).
If the bond output is \(\xi\) with one-site supports \(X,Y\)
(the spans of its left and right Schmidt vectors), this image is
\(\mathbb C\xi\otimes W\). On sites 2 and 3, \(F_{23}\) must preserve
\(Y\otimes W\), since the next factor \(F_{13}\) cannot change the
support on site 2. On sites 1 and 3, the support just before \(F_{13}\)
is contained in \(X\otimes W\), and just after it equals \(X\otimes W\).
Invertibility forces equality of the two dimensions, so \(F\) also
preserves \(X\otimes W\). The second mixed commutator
\([h_{23},F_{13}F_{12}]=0\) gives the same statements with the slots
reversed. In particular, for a product output \(F\) fixes the output
tensor line. Applying the argument on the dual space to \(h^{\mathsf T}\)
and \(F^{-\mathsf T}\) gives preservation of the input radicals in both
slots. They remain the radicals of the transformed input because the inverse
twist gives the reverse inclusion, by the Yang--Baxter relation for
\(F\). Thus \(F\) acts by a nonzero scalar on each tensor product of
two marked lines in the rank-two test above, preserving its zero
pairings. For a product input the dual argument fixes the input factor
lines as well. The same signatures therefore separate the 17 classes
under these additional twists.


\begin{thebibliography}{99}
\bibitem{Shattering}
V.~Khemani, M.~Hermele and R.~Nandkishore, Localization from Hilbert
space shattering: From theory to physical realizations,
\href{https://doi.org/10.1103/PhysRevB.101.174204}{Phys. Rev. B
\textbf{101}, 174204 (2020)}.
\bibitem{SalaDipole}
P.~Sala, T.~Rakovszky, R.~Verresen, M.~Knap and F.~Pollmann,
Ergodicity Breaking Arising from Hilbert Space Fragmentation in
Dipole-Conserving Hamiltonians,
\href{https://doi.org/10.1103/PhysRevX.10.011047}{Phys. Rev. X
\textbf{10}, 011047 (2020)}.
\bibitem{HSFReview}
S.~Moudgalya, B.~A.~Bernevig and N.~Regnault, Quantum many-body scars
and Hilbert space fragmentation: a review of exact results,
\href{https://doi.org/10.1088/1361-6633/ac73a0}{Rep. Prog. Phys.
\textbf{85}, 086501 (2022)}.
\bibitem{DimerSectors}
G.~I.~Menon, M.~Barma and D.~Dhar, Conservation laws and integrability
of a one-dimensional model of diffusing dimers,
\href{https://doi.org/10.1007/BF02183622}{J. Stat. Phys. \textbf{86},
1237--1263 (1997)}.
\bibitem{PairKrylov}
S.~Moudgalya, A.~Prem, N.~Regnault, R.~Nandkishore and B.~A.~Bernevig,
Thermalization and its absence within Krylov subspaces of a constrained
Hamiltonian, in \emph{Memorial Volume for Shoucheng Zhang}
(World Scientific, 2021),
\href{https://doi.org/10.1142/9789811231711_0009}{pp.~147--209}.
\bibitem{TiltedFermi}
T.~Kohlert, S.~Scherg, P.~Sala, F.~Pollmann, B.~Hebbe Madhusudhana,
I.~Bloch and M.~Aidelsburger, Exploring the regime of fragmentation in
strongly tilted Fermi--Hubbard chains,
\href{https://doi.org/10.1103/PhysRevLett.130.010201}{Phys. Rev. Lett.
\textbf{130}, 010201 (2023)}.
\bibitem{TiltedBose}
D.~Adler, D.~Wei, M.~Will, K.~Srakaew, S.~Agrawal, P.~Weckesser,
R.~Moessner, F.~Pollmann, I.~Bloch and J.~Zeiher, Observation of
Hilbert space fragmentation and fractonic excitations in 2D,
\href{https://doi.org/10.1038/s41586-024-08188-0}{Nature \textbf{636},
80--85 (2024)}.
\bibitem{FoldedSpinChain}
B.~Pozsgay, T.~Gombor, A.~Hutsalyuk, Y.~Jiang, L.~Pristy\'ak and
E.~Vernier, Integrable spin chain with Hilbert space fragmentation and
solvable real-time dynamics,
\href{https://doi.org/10.1103/PhysRevE.104.044106}{Phys. Rev. E
\textbf{104}, 044106 (2021)}.
\bibitem{GhoshFragmentation}
S.~Ghosh, K.~Sengupta and I.~Paul, Hilbert space fragmentation imposed
real spectrum of non-Hermitian systems,
\href{https://doi.org/10.1103/PhysRevB.109.045145}{Phys. Rev. B
\textbf{109}, 045145 (2024)}.
\bibitem{MM}
S.~Moudgalya and O.~I.~Motrunich, Hilbert Space Fragmentation and
Commutant Algebras, \href{https://doi.org/10.1103/PhysRevX.12.011050}{Phys.
Rev. X \textbf{12}, 011050 (2022)}.
\bibitem{MPK}
S.~Maity, P.~Padmanabhan and V.~Korepin, Non-hermitian integrable systems
from constant non-invertible solutions of the Yang--Baxter equation,
\href{https://doi.org/10.1007/JHEP05(2025)206}{J. High Energy Phys.
\textbf{05} (2025) 206}, \href{https://arxiv.org/abs/2503.08109}{arXiv:2503.08109}.
\bibitem{PK}
P.~Padmanabhan and V.~Korepin, A Universal Algebraic Construction of
Hilbert-Space Fragmentation, \href{https://arxiv.org/abs/2611.19470}{arXiv:2611.19470 (2026)}.
\bibitem{Hietarinta}
J.~Hietarinta, Solving the two-dimensional constant quantum Yang--Baxter
equation, \href{https://doi.org/10.1063/1.530185}{J. Math. Phys.
\textbf{34}, 1725--1756 (1993)}.
\bibitem{SF78}
E.~K.~Sklyanin and L.~D.~Faddeev, Quantum mechanical approach to
completely integrable field theory models,
\href{https://www.mathnet.ru/eng/dan42231}{Dokl. Akad. Nauk SSSR
\textbf{243}, 1430--1433 (1978)} [Sov. Phys. Dokl. \textbf{23},
902--904 (1978)].
\bibitem{STF79}
E.~K.~Sklyanin, L.~A.~Takhtajan and L.~D.~Faddeev, Quantum inverse
problem method. I, \href{https://doi.org/10.1007/BF01018718}
{Theor. Math. Phys. \textbf{40}, 688--706 (1979)}.
\bibitem{IK}
A.~G.~Izergin and V.~E.~Korepin, The quantum inverse scattering method
approach to correlation functions, \href{https://doi.org/10.1007/BF01212350}
{Commun. Math. Phys. \textbf{94}, 67--92 (1984)}.
\bibitem{KBI}
V.~E.~Korepin, N.~M.~Bogoliubov and A.~G.~Izergin,
\href{https://doi.org/10.1017/CBO9780511628832}{\emph{Quantum Inverse
Scattering Method and Correlation Functions}} (Cambridge University Press,
Cambridge, 1993), Part~II, Chaps.~VI--VII.
\bibitem{SlavnovReview}
N.~A.~Slavnov, The algebraic Bethe ansatz and quantum integrable systems,
\href{https://doi.org/10.1070/RM2007v062n04ABEH004430}{Russ. Math. Surv.
\textbf{62}, 727--766 (2007)}.
\bibitem{SlavnovLectures}
N.~A.~Slavnov, Algebraic Bethe ansatz,
\href{https://arxiv.org/abs/1804.07350}{arXiv:1804.07350 (2018)}.
\bibitem{SlavnovBook}
N.~A.~Slavnov, \href{https://doi.org/10.1142/12776}{\emph{Algebraic Bethe
Ansatz and Correlation Functions: An Advanced Course}} (World Scientific,
Singapore, 2022).
\bibitem{Cherednik}
I.~V.~Cherednik, Factorizing particles on a half-line and root systems,
\href{https://doi.org/10.1007/BF01038545}
{Theor. Math. Phys. \textbf{61}, 977--983 (1984)}.
\bibitem{Sklyanin}
E.~K.~Sklyanin, Boundary conditions for integrable quantum systems,
\href{https://doi.org/10.1088/0305-4470/21/10/015}{J. Phys. A
\textbf{21}, 2375--2389 (1988)}.
\bibitem{Vlaar}
B.~Vlaar, Boundary transfer matrices and boundary quantum KZ equations,
\href{https://doi.org/10.1063/1.4927305}{J. Math. Phys. \textbf{56},
071705 (2015)}.
\bibitem{KulishSklyanin}
P.~P.~Kulish and E.~K.~Sklyanin, Algebraic structures related to
reflection equations, \href{https://doi.org/10.1088/0305-4470/25/22/022}
{J. Phys. A \textbf{25}, 5963 (1992)}.
\bibitem{StanleyEC}
R.~P.~Stanley, \href{https://doi.org/10.1017/CBO9781139058520}
{\emph{Enumerative Combinatorics}}, Vol.~1, 2nd ed.
(Cambridge University Press, Cambridge, 2012).
\bibitem{GYBEsym}
P.~Padmanabhan, S.~Maity, A.~Sinha and V.~Korepin, Symmetries of the
Generalized Yang--Baxter Equations,
\href{https://doi.org/10.46298/ocnmp.18633}
{Open Commun. Nonlinear Math. Phys., Special Issue in Honour of
J.~Hietarinta, 251--281 (2026)}, \href{https://arxiv.org/abs/2606.26510}{arXiv:2606.26510}.
\bibitem{FlajoletSedgewick}
P.~Flajolet and R.~Sedgewick, \href{https://doi.org/10.1017/CBO9780511801655}
{\emph{Analytic Combinatorics}} (Cambridge University Press, Cambridge, 2009).
\bibitem{FK}
S.~Fomin and A.~N.~Kirillov, The Yang--Baxter equation, symmetric functions,
and Schubert polynomials, \href{https://doi.org/10.1016/0012-365X(95)00132-G}
{Discrete Math. \textbf{153}, 123--143 (1996)}.
\bibitem{QutritStudy}
S.~Maity, P.~Padmanabhan and V.~Korepin, Yang--Baxter integrability and
Hilbert-space fragmentation in non-Hermitian qutrit chains, {\it In preparation}.
\bibitem{BellonMaillardViallet}
M.~P.~Bellon, J.-M.~Maillard and C.~Viallet, Infinite discrete symmetry
group for the Yang--Baxter equations. Vertex models,
\href{https://doi.org/10.1016/0370-2693(91)90974-U}
{Phys. Lett. B \textbf{260}, 87--100 (1991)}.
\bibitem{ReshetikhinTwist}
N.~Reshetikhin, Multiparameter quantum groups and twisted
quasitriangular Hopf algebras,
\href{https://doi.org/10.1007/BF00626530}
{Lett. Math. Phys. \textbf{20}, 331--335 (1990)}.
\bibitem{ScaleneHidden}
P.~Padmanabhan, S.~Maity and V.~Korepin,
Scalene Yang--Baxter triples as a source of hidden symmetries beyond
the ordinary Yang--Baxter equation,
\href{https://arxiv.org/abs/2608.09081}{arXiv:2608.09081} (2026).
\bibitem{ScaleneAffine}
P.~Padmanabhan, S.~Maity and V.~Korepin,
Multiparameter quantum affine spaces and the scalene Yang--Baxter
equation,
\href{https://doi.org/10.1016/j.nuclphysb.2026.117707}
{Nucl. Phys. B \textbf{1033}, 117707 (2026)}.
\bibitem{ScaleneAdditive}
P.~Padmanabhan, S.~Maity, A.~Sinha and V.~Korepin,
Scalene Yang--Baxter triples realize all nearest-neighbour Hamiltonians
with global additive symmetries,
\href{https://arxiv.org/abs/2610.13782}{arXiv:2610.13782} (2026).
\end{thebibliography}
\end{document}